\documentclass[review]{elsarticle}

\usepackage{chngcntr}
\usepackage{multirow}
\usepackage{lscape}
\usepackage{color}
\usepackage{caption}
\usepackage{subcaption}
\usepackage[textwidth=340pt,textheight=550pt]{geometry}
\usepackage[colorlinks]{hyperref}

\journal{Applied Energy}

\begin{document}
	
	\begin{frontmatter}
		
		\title{\textcolor{black}{Short-term CO$_2$ emissions forecasting based on decomposition approaches and its impact on electricity market scheduling}}
		
		
		\author[mymainaddress]{Neeraj Bokde}
		
		\author[ento]{Bo Tranberg}
		
		\author[mymainaddress]{Gorm Bruun Andresen\corref{mycorrespondingauthor}}
		\cortext[mycorrespondingauthor]{Corresponding author}
		\ead{gba@eng.au.dk}
		
		\address[mymainaddress]{Department of Engineering - Renewable Energy and Thermodynamics, Aarhus University, 8000, Denmark}
		\address[ento]{Ento Labs ApS, Inge Lehmanns Gade 10, 6, 8000, Aarhus C, Denmark}
		
		\begin{abstract}
				The world is facing major challenges related to global warming and emissions of greenhouse gases is a major causing factor. In 2017, energy industries accounted for 46\% of all CO$_2$ emissions globally, which shows a large potential for reduction. This paper proposes a novel short-term CO$_2$ emissions forecast to enable intelligent scheduling of flexible electricity consumption to minimize the resulting CO$_2$ emissions. Two proposed time series decomposition methods are developed for short-term forecasting of the CO$_2$ emissions of electricity. These are in turn bench-marked against a set of state-of-the-art models. The result is a new forecasting method with a 48-hour horizon targeted the day-ahead electricity market. Forecasting benchmarks for France show that the new method has a mean absolute percentage error that is 25\% lower than the best performing state-of-the-art model. Further, application of the forecast for scheduling flexible electricity consumption is studied for five European countries. Scheduling a flexible block of 4 hours of electricity consumption in a 24 hour interval can on average reduce the resulting CO2 emissions by 25\% in France, 17\% in Germany, 69\% in Norway, 20\% in Denmark, and just 3\% in Poland when compared to consuming at random intervals during the day.
		\end{abstract}
		
		
	\end{frontmatter}
	
	
	\section{Introduction}
	The modern human society is facing many challenges related to global warming and emissions of greenhouse gases into the atmosphere is a major factor responsible for it \cite{IPCC2018, XX2}.

	According to a report by the European Commission, in 2017 China was responsible for 29\% of total CO$_2$ emission and surpassed other economies including the USA, Europe, and India which contributed 14\%, 10\%, and 7\%, respectively \cite{EUemissions}. Further, this report revealed that the global annual CO$_2$ emissions and global per capita emissions have climbed to 37 billion metric tons and 4.9 metric tons in the year 2014, respectively. In the same year, energy industries accounted for 46\% of all CO$_2$ emissions globally \cite{IEAemissions}. This shows a large potential for reduction.

	In 2009, most countries have shown their intentions to reduce CO$_2$ emission at the Copenhagen Global Climate Conference (also known as the 2009 United Nations Climate Change Conference) \cite{XX5}. During this conference, the European Union and the USA committed to lowering the CO$_2$ emission by 20 to 30\% and 4\% of the level in 1990, respectively. Whereas, by 2020, the developing economies, China and India targeted reducing the CO$_2$ emissions per GDP by around 40\% and 25\% of the level in 2005 by 2020, respectively. Furthermore, most countries as a united globe have set the targets for upcoming decades to reduce greenhouse gases significantly \cite{XX6}. In addition to this, the Paris agreement (famously named as 20/20/20) set to reduce CO$_2$ emissions by 20\%, increase the renewable energy share by 20\%, and increase energy efficiency by 20\% for the whole World \cite{XX7}. The country-wise profiles related to CO$_2$ emissions and climate change are discussed in detail by the CarbonBrief team \cite{XX8}.
	
	During recent years, most countries targeted specific CO$_2$ emissions to meet global warming challenges. This leads to lots of research contributions on the trends of CO$_2$ emissions to assist energy planning policies. Although, there are numerous techniques proposed by research teams throughout the World. Ye et. al. \cite{XX9} roughly categorized these techniques into two forms, namely, target decomposition and policy evaluation. Target decomposition is a simple and easy technique, which decomposes the targeted goals of CO$_2$ emissions into sub-divisions. These sub-divisions can be based on various criteria based on different departments or provinces \cite{XX10, XX11}. However, these studies quantify the potential of a specific policy. Whereas, in policy evaluation techniques, policies to reduce CO$_2$ emissions are evaluated under present scenarios. These techniques include statistical, econometric and computer simulation approaches to forecast CO$_2$ emissions.

	
		Forecasting CO$_2$ intensity and related CO$_2$ emissions is important today and will become urgent in the future to recommend effective CO$_2$ emissions reduction policies to the policymakers. Furthermore, it may aid the system in integrating renewable energy. Most countries are now under tremendous pressure to limit reduce their CO$_2$ emissions. Hence, it is of great importance to explore and analyze the responsible factors for CO$_2$ emissions. Accurate forecasting models enable such analyses.
		
		Throughout Europe, there are multiple bidding areas in many countries in the day-ahead electricity market. For each bidding area, the transmission capacity availabilities vary and this may congest the power flow between neighboring bidding areas. Such congestion leads to differences in the price between bidding areas. Among the energy markets in Europe are Nord Pool covering Scandinavian countries \cite{m2}, and EPEX SPOT covering North-West Europe \cite{m1}.
		
		The purpose of electricity markets is price formation to establish equilibrium between supply and demand. This is done on an hourly level in the day-ahead market for electricity. The day-ahead market clears at noon and provides an hourly price for the 24 hours of the following day based on matching bids and offers received from producers and consumers. When bidding, one needs to specify the amount of energy in MWh for an hour to be bought (consumed) or sold (produced) at a certain price level (Eur/MWh) for the next day.
		
\begin{figure}[h]
			\centering{\includegraphics[width=5in]{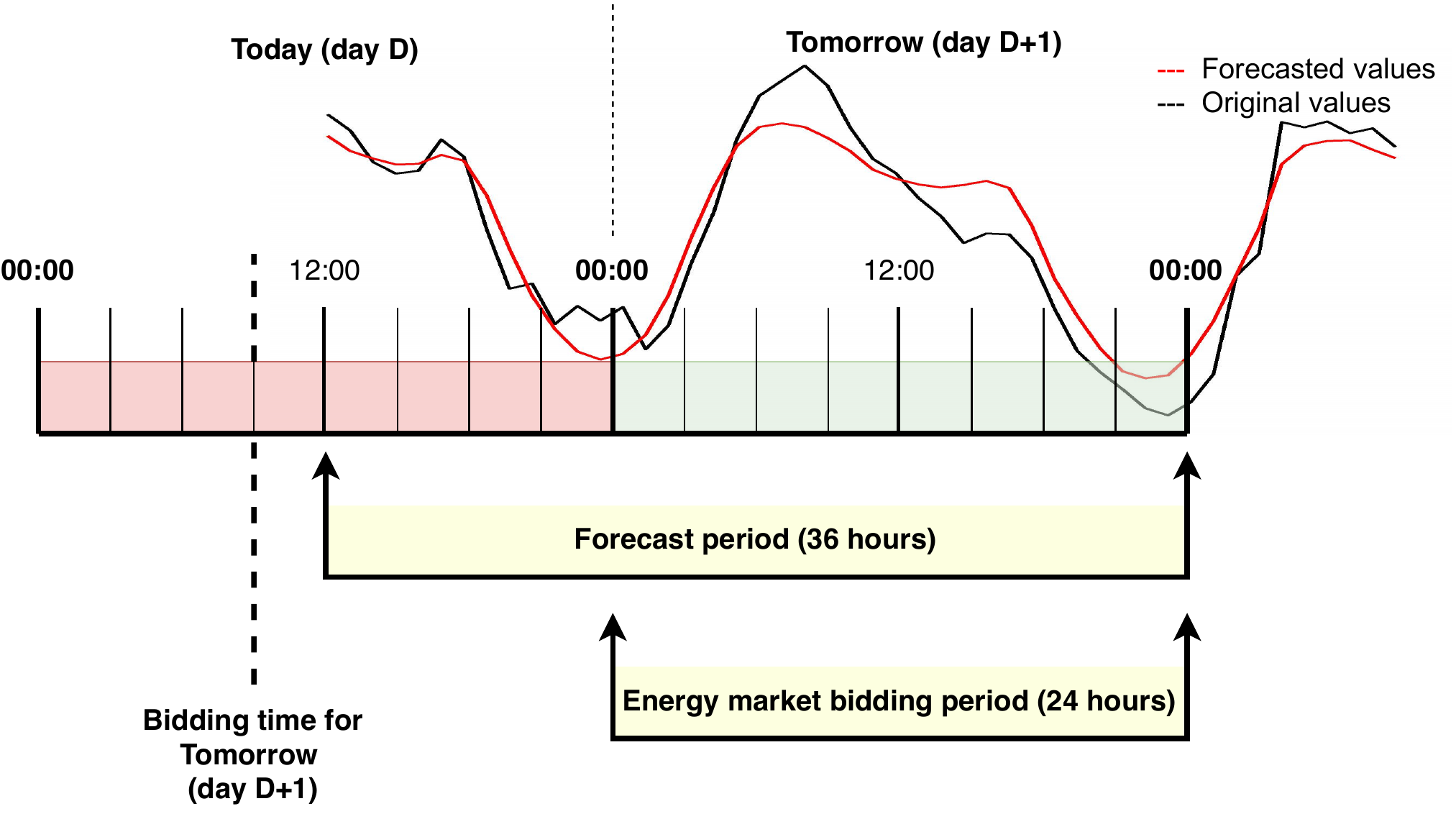}}
			\caption{Time framework to forecast for the day-ahead energy market.}
			\label{fig_sc}
		\end{figure}

		Consider the time framework as shown in Figure \ref{fig_sc} for further understanding of the electricity market bidding. So far, such time frameworks have been used to plan the market price forecast in the day-ahead energy market. In Figure \ref{fig_sc}, the day ($D+1$), i.e. tomorrow, is the forecasting period of interest and for this auction, the consumer needs to bid before 12.00 AM today, i.e. day $D$. In addition, considering some buffer periods and time to place bids in the market, it will be better for the consumer to be ready with such a forecast analysis at 09:00 - 10.00 AM on the day $D$. Hence, this day ahead electricity market needs 12-36 hours, hourly resolved forecasts.
				
		Normally, consumers schedule the time slots with reference to electricity prices and it is based on the electricity prices forecasts. The motivation is to reduce the overall cost of consumed electricity for the desired scheduled period \cite{voronin2013price}.
		
		In recent years many countries, public buildings, public infrastructures, and institutes {started} to take initiatives to understand the sources and the nature of the energy they are consuming. Such entities want to keep track of their CO$_2$ emissions. In relation to energy consumption, this means focusing on greener energy with lower CO$_2$ intensity. In the case of bidding in the day-ahead market, the consumer could schedule not just after the forecasted price, but also accordingly to forecasted CO$_2$ intensity as a primary or secondary objective. However, this can only be achieved if appropriate forecasts of the CO$_2$ is available. With this motivation in mind, this study investigates new, appropriate models for electricity bidding-friendly multi-step hours CO$_2$ emissions forecasts.
	
	
	
		From a system perspective, forecasts like the one proposed in this study are important for the scheduling of flexible electricity consumption to have an effect on the increasing integration of renewable energy resources in the electricity system. Flexible consumers must be able to place bids in the day-ahead market to provide timely signals for the market participants e.g., to adjust production plans accordingly.

	In this article, two time series decomposition methods are compared for short-term forecasting of the CO$_2$ emissions of electricity. First, the CO$_2$ intensities are obtained using the power flow tracing technique presented in \cite{Tranberg2015} and combined with country-specific average CO$_2$ intensity per generation technology as proposed in \cite{tranberg2019real}. Second, these intensity time series are decomposed into three components with two approaches. Finally, three different forecasting models are employed to predict all components, and these predicted magnitudes are combined to obtain the final forecasted values for a given CO$_2$ intensity series. The performance of both decomposition approaches is then compared with several state-of-the-art forecasting methods. The CO$_2$ intensity series exhibit non-stationary and nonlinear patterns, which are forecasted effectively and precisely with the multi-scale integrated prediction methods.
	
	Related studies in the literature focus on long-term forecasts of CO$_2$ emissions. For electricity consumers to be able to provide flexible demand to the system they need more accurate short-term forecasts as the novel methods we are proposing in this study. Though these new methods are short-term forecasts, they predict values that are 36 to 48 hours ahead in time in a single step, which is a novel approach for this application.
	\textcolor{black}{Consequently, this paper contributes to the literature as follows:
	\begin{itemize}
		\item The models to forecast the short-term CO$_2$ intensity time-series for 36–48 hours ahead values, which are necessary for electricity market bidding.
		\item Discussing the impact of forecasted CO$_2$ emissions on the electricity market schedule.
		\item Enabling flexible electricity consumption scheduling to minimize CO$_2$ emissions.
	\end{itemize}}
	
	The remainder of this article is organized as follows: Section 2 discusses the motivation behind 48 steps ahead forecasting of CO$_2$ intensity and its effects on electricity markets. Section 3 provides a precise review of various models employed for CO$_2$ intensity forecasting. The proposed methodology along with two decomposition strategies are discussed in Section 4. Further, the performance evaluation of the proposed methods is studied on sample standard data sets in Section 5. In Section 6, a case study is discussed which evaluates the performance of the proposed methods for short-term multi-step CO$_2$ intensity data set forecasting. Section 7 presents a case study on potential CO$_2$ emissions reductions resulting from scheduling flexible electricity consumption. Finally, the conclusions of the paper are summarized in Section 8. All abbreviations and variables used throughout the paper are listed in Table~\ref{TA}.
	
	\begin{table}[]
		\centering
		\caption{Nomenclature.}
		\label{TA} 
		\makebox[\columnwidth]{
			\begin{tabular}{|ll|ll|}
				\hline
				\textbf{Abbreviations} &  & \textbf{Variables} & \\
				ARIMA & AutoRegressive Integrated Moving Average & $x(t)$ & a time series \\
				CO2  & Carbon Dioxide         & $C_i(t)$ & $i^{th}$ IMF \\
				DK & Denmark & $r_m(t)$ & the final residual \\
				DPSF & Differenced Pattern Sequence-based & $e_L(t)$ & Lower envelope \\
				& Forecasting method & $e_H(t)$ & Upper envelope \\
				EEMD & Ensemble Empirical Mode Decomposition & $a(t)$ & Average of lower\\
				EMD &  Empirical Mode Decomposition & & and upper evelopes\\
				ENTSO-E & European Network of Transmission  & $M$ & Ensemble numbers \\
				& System Operators for Electricity & $e$ & Amplitude of the \\
				EPEX SPOT & European Power Exchange  & & white noise series \\
				FFNN & Feedforward Neural Network  & $e^{'}$ & Final standard \\
				GDP & Gross Domestic Product & & deviation of error \\
				HVAC & Heating, Ventilation, and & $\mathrm{X_{i}}$ & Observed data \\ 
				& Air Conditioning  & $\mathrm{\hat{X}_{i}}$ & Forecasted data \\
				IMF & Intrinsic Mode Functions & $\mathrm{N}$ & Number of time steps\\ 
				IPT & Integrated Powertrain & &  for forecast evaluation \\
				kWh  & Kilo-Watt Hour & & \\
				LSSVM & Least-squares support-vector machine & & \\
				LSTM & Long-Short-Term Memory & & \\
				MAE  & Mean Absolute Error    & & \\
				MAPE & Mean Absolute Percentage Error & & \\ 
				max. LE & Maximum Lyapunov Exponent & & \\
				MWh & Mega-Watt Hour & & \\
				NWP & Numerical Weather Prediction  & & \\
				PSF & Pattern Sequence-based Forecasting method & &  \\
				RMSE & Root Mean Square Error & &  \\
				SOPR & Second-Order Polynomial Regression & & \\
				SVM & Support Vector Machine & & \\		
				\hline
			\end{tabular} 
		}
	\end{table} 

	\section{Method}
	
	
		\subsection{Review of CO$_2$ emissions forecasts}
		Long-term CO$_2$ emissions forecasting enables policymakers and planners to estimate and reduce the future carbon footprint. However, the short-term CO$_2$ emissions forecasting is critical to understand how much CO$_2$ will be emitted so that e.g., producers and consumers of electricity can schedule their production and consumption to minimize their resulting emissions.
		
		Globally, long-term CO$_2$ emissions have been an attractive topic for research and discussion, but there are very few articles discussing short-term forecasting. Finenko et. al. \cite{XX28} is one of the initial articles which analyzed the hourly patterns of CO$_2$ emissions for electricity generation and stated that most CO$_2$ emissions reduction can be achieved during day time with a case study of Singapore. Mason et. al. \cite{XX29} proposed a case study in Ireland, which successfully forecasts short-term CO$_2$ emissions using evolutionary neural networks. There are few case studies \cite{XX30, XX31, XX32} which predicted air pollutants and ozone concentrations for short time horizons, which were closely related to carbon footprints. Similarly, a light-use-efficiency model based on numerical weather prediction (NWP) method is proposed for short-term forecasting of CO$_2$ fluxes in Europe \cite{XXN3}.
		
		For the first time, grid CO$_2$ intensity in support of heating, ventilation, and air-conditioning (HVAC) load is forecasted for the day ahead horizon by \cite{XX33}. This study foresees the opportune time of the day for carbon saving based on CO$_2$ intensity forecasts and accordingly shutdown of the HVAC plant.

	
	In contrast to the CO$_2$ emissions time series used for long-term forecasting, the one used for short-term forecasting is of higher complexity. For short-term forecasting, the data set used has an hourly resolution, hence it exhibits rapidly changing patterns. This highly chaotic time series is challenging to forecast with the standard set of forecasting methods. As discussed in the literature review, the models used for long-term CO$_2$ emissions forecasting were mainly focused on understanding and estimating the trends of the time series. However, in time series used for short-term forecasting, there are many components apart from trends, which need to be considered. Generally, a series can be segmented into three parts; these are the seasonal, trend and random components. The seasonal component shows the patterns in the series that repeat with a fixed period, whereas the trend component shows the tendency of a series to increase or decrease over a long period of time, which is a smoother, general and long-term tendency. Finally, the random component is purely irregular in nature. It does not represent any pattern and hence is an uncontrollable and unpredictable part of the series. Minimizing the share of the random component, the time series becomes more informative and non-chaotic in nature. Since all these components individually have different patterns, it is not adequate to handle them jointly with a single forecasting model as it is important to analyze the behavior of each component individually.
	
	In the present study, the time series is decomposed into the seasonal, trend and random component series with two different approaches, and these components are further forecasted with different models. Eventually, forecasts of CO$_2$ intensity are achieved and a comparison of the mentioned decomposition approaches is done.
	
	\subsection{Decomposition with first approach}
	
\begin{figure}[]
		\centering
		\begin{subfigure}[b]{\textwidth}
			\centering{\includegraphics[width=4.5in]{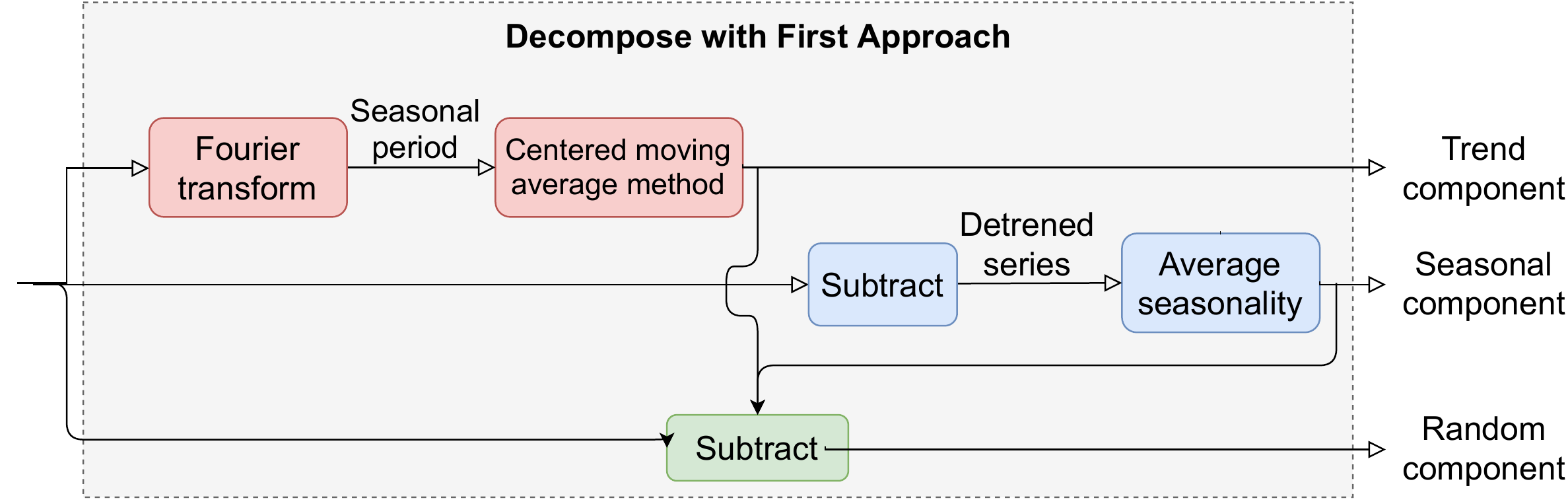}}
			\caption{Decomposition with moving averages.}
			\label{fig4}
		\end{subfigure}
		\par\bigskip
		\begin{subfigure}[b]{\textwidth}
			\centering{\includegraphics[width=4.5in]{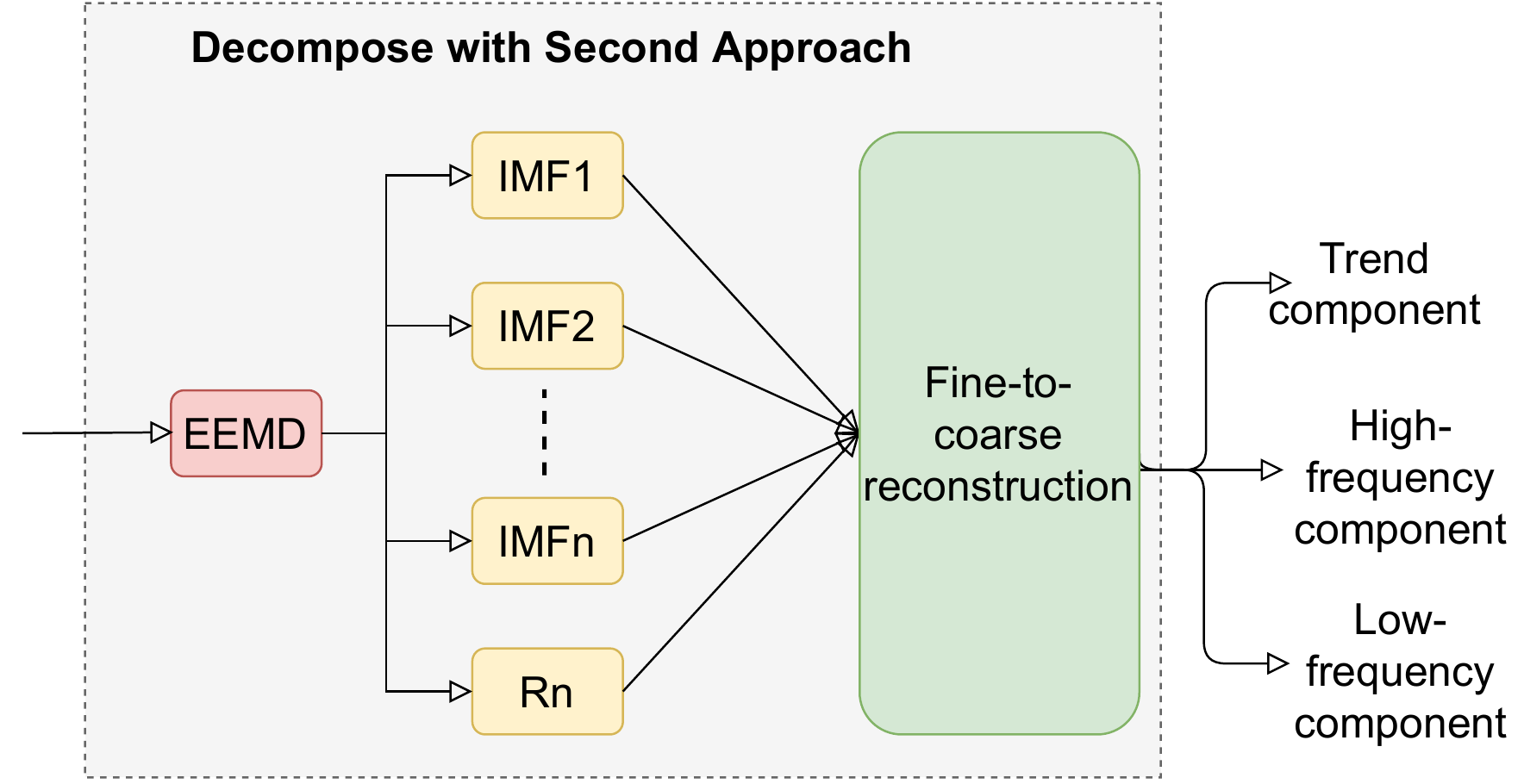}}
			\caption{Decomposition with EEMD method.}
			\label{fig5}
		\end{subfigure}
		\caption{Block diagram of the (a) moving averages and (b) EEMD method based decomposition approaches.}
		\label{fig:th graphs}
	\end{figure}

	The first approach is a basic and well-known methodology, which first derives the trend, seasonal and noise components one by one with various statistical steps. The first step is to detect the trend in the time series using the centered moving average method. In this step, a moving window with the length of the seasonal period of the time series is used. The seasonality of a time series is either observed by a seasonal period of the time series (daily, weekly, monthly, or yearly) or estimated using the Fourier transform method. Then, this moving window is replaced with the average of the window and the trend series is obtained. The second step is detrending the time series by removing the trend series obtained in the first step from the original time series. The detrended series appears to be seasonal but infected with a noise component. Now, to determine the actual seasonality component, the average seasonality needs to be calculated. It is obtained in the third step by adding seasonality of the detrended series together and dividing it by the seasonality period. In the final step, the random noise component is extracted by subtracting seasonal and trend components from the original time series. The block-diagram of this approach is as shown in Figure \ref{fig4}.
	
	\subsection{Decomposition with second approach}
	
	Furthermore, the second approach is based on a decomposition methodology. This ensemble empirical mode decomposition (EEMD) method is used to decompose the time series into a number of sub-series. It is a modification of the EMD method, which is an empirical and effective decomposition method with a less complex procedure. EMD is efficient at handling non-linear and non-stationary time series.
	The methodology was introduced by \cite{XX34} and since then, it has become increasingly popular in the research field of forecasting, and substantially enhancing prediction process \cite{XX35}. With the methodology, the time series is decomposed into numerous intrinsic mode functions (IMF) and one residual. The process is termed as the EMD sifting process and can be expressed linearly as:
	
	\begin{equation}
	\label{e1}
	x(t) = \sum_{i=1}^{m}C_{i}(t) + r_{m}(t)\
	\end{equation}
	where, $m$ is the number of IMFs, $C_i(t)$ is the $i$th IMF at the time $t$, and $r_m(t)$ is the final residual of the time series $x(t)$.  All $C_i(t)$ have zero means and are nearly orthogonal to each other \cite{XX36}.
	
	In the processes, firstly all local minima and maxima values of $x(t)$ are identified. Then the cubic spline interpolation is used to obtain the lower envelope $e_L(t)$ and upper envelop $e_H(t)$ for the minima and maxima values. The average value of $a(t)$ of the lower and upper envelope is calculated as:
	
	\begin{equation}
	\label{e2}
	a(t)= \frac{(e_L (t)+ e_H (t))}{2}.
	\end{equation}
	
	The first component $d(t)$ is given by:
	
	\begin{equation}
	\label{e3}
	d(t)=x(t)-a(t).
	\end{equation}
	
	After the above step, if $d(t)$ meets the IMF properties, then it is considered as the first IMF. Otherwise, $x(t)$ is replaced by $d(t)$, and the above process is repeated. In case of the residual function $r(t)$, when it becomes monotonic or only extreme values are present from which no more IMF is achieved, the process is terminated \cite{XX37}. The IMF should satisfy the following two condition:
	
	\begin{itemize}
		\item The number of extrema and zero crossings of the data set must either be equal or differ at most by one.
		\item At any point, the mean value defined by $e_L(t)$ and $e_H(t)$ should be zero.
	\end{itemize}

	The EEMD method is an adaptive approach developed from EMD. It was proposed by \cite{XX38}. The methodology has an improvement over EMD, as it adds finite white noise to the original data, to overcome the problem of mode mixing that exists with EMD. In the mode mixing problem, the data may not decompose properly and hence a similar scale series are found to exist in different IMFs \cite{XX36}, which results in loss of physical meaning in them. To decompose the time series by EEMD the following steps have been adopted \cite{XX38}: Random white noise having zero mean and given standard deviation $e$ has to be added to the available time series $x(t)$.
	
	This time series is decomposed into various IMFs and one residual component. The above procedure is repeated with different white noise series each time. The mean of corresponding IMFs of the decomposition is obtained as the final results. The addition of all the IMFs will not be the same at the end \cite{bokde2018analysis}. The effect of white noise can be minimized based on the following equation \cite{XX38} and \cite{XX40}.
	
	\begin{equation}
	\label{e4}
	e^{'}= \frac{e}{\sqrt M},
	\end{equation}
	where $M$ is the ensemble numbers, $e$ is the amplitude of the added white noise series, and $e^{'}$ is the final standard deviation of error. Many researchers have supported the use of EEMD for the decomposition of time series and it has been implemented successfully in various fields \cite{bokde2019review}.
	
	The case studies in \cite{XX41} and \cite{XX42} applied EEMD to time series of carbon and crude oil prices and decomposed them into three IMF components: high frequency, low frequency, and trend as shown in Figure \ref{fig5}. These components seem equivalent, but an alternative to the seasonality, noise, and trend components, respectively. A fine-to-coarse reconstruction algorithm \cite{XX42} is used to separate the IMFs into high- and low-frequency components, whereas the residual is the trend component. In this procedure, the mean of the algorithm is traced and when it departs significantly from zero for the `$n$'th IMF, the partial reconstruction using IMF1-IMF($n-1$) represents the high-frequency component. Whereas IMF$n$ to the last IMF comprises the low-frequency component. The detailed analysis is further discussed in the following sections for the proposed case study.
	
	
	
	\subsection{Significance of decomposition in forecast}
	
	The significance of decomposition methods in a forecasting model can be explained with human experience and observations. Consider an example of weather in a particular region. If one observed extreme cold in December and excessive rainfall in August every year, this suggests that next year, there will be another colder December as well as sufficient rain in August.
	Similarly, the decomposed seasonal component captures the seasonal patterns which repeat after a specific time interval and a forecasting model can easily observe and estimate the future within such seasonal or high-frequency component series.
	
	Again, though it is expected, extreme cold in December sometimes is delayed or the extreme cold peak is not occurring in a particular year. Besides, sometimes, unexpected rainfall is observed in the winter season. All these unexpected observations are nothing but the random activities based on various natural or unnatural processes of the Earth. The random or low-frequency components obtained with the discussed decomposition approaches represent this random behavior of the time series.
	
\begin{figure}[h]
		\centering{\includegraphics[width=\textwidth]{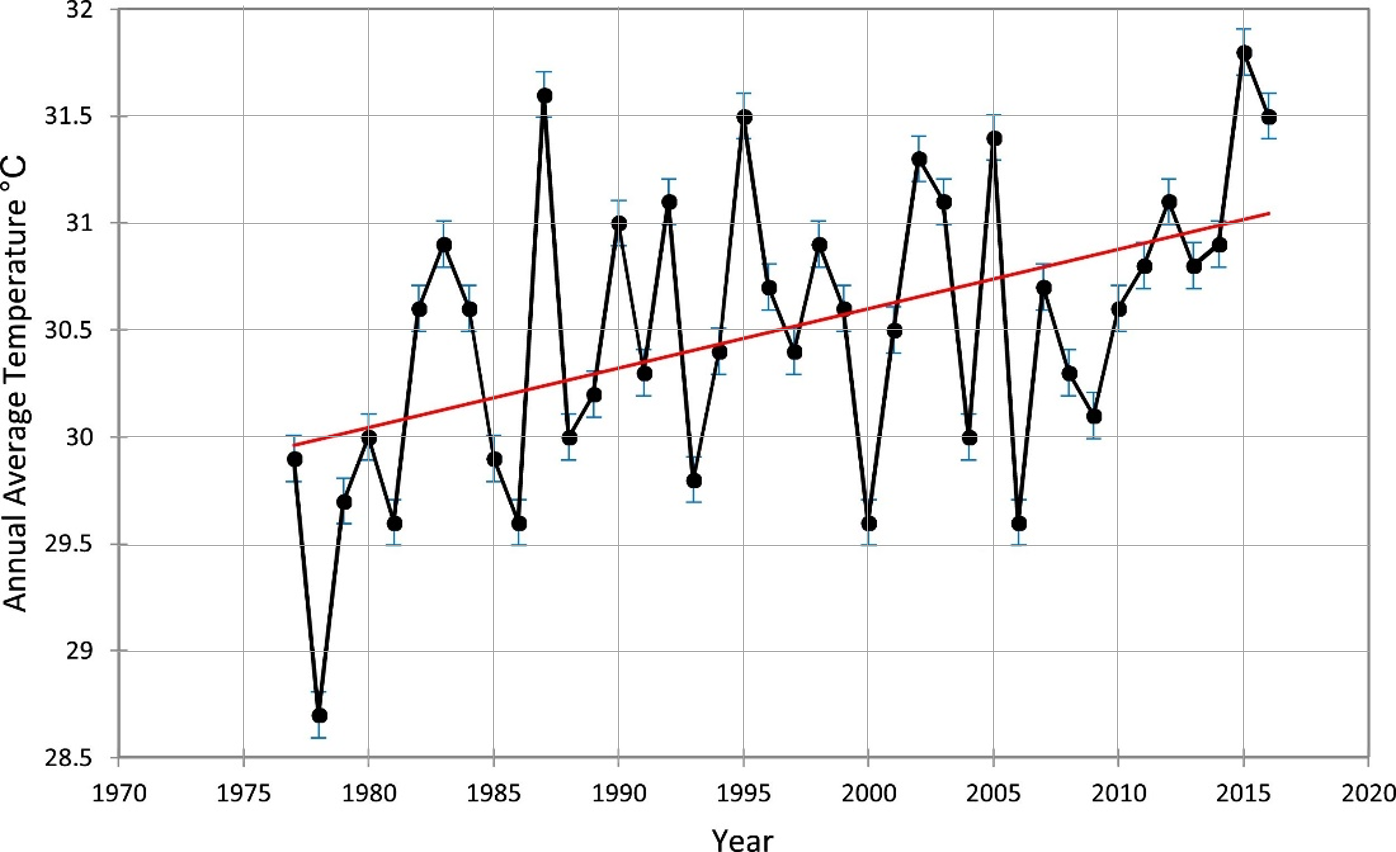}}
		\caption{A trend showing annual average temperature for years 1977–2016 (Source: \cite{dube2018climate})}
		\label{trend}
	\end{figure}

	Further, climate change is another example. Earlier studies \cite{barnard2017extreme, dube2018climate} confirms significant increase in temperature during decades as shown in Figure \ref{trend}. The broad sense of temperature change in the century represented a trend line. Again, these trend series can help a forecasting model to estimate such long-term behavior present in the time series. Both discussed decomposition approaches generate trend components with different strategies.
	
	In this way, a proper strategy to forecast each of the series (seasonal, trend, and random components) individually with distinct suitable models may better reproduce the respective patterns in the forecasted series. Eventually, a more accurate forecast can be achieved.

	\subsection{Selection of forecasting models}
	
\begin{figure}[h]
		\centering{\includegraphics[width=\textwidth]{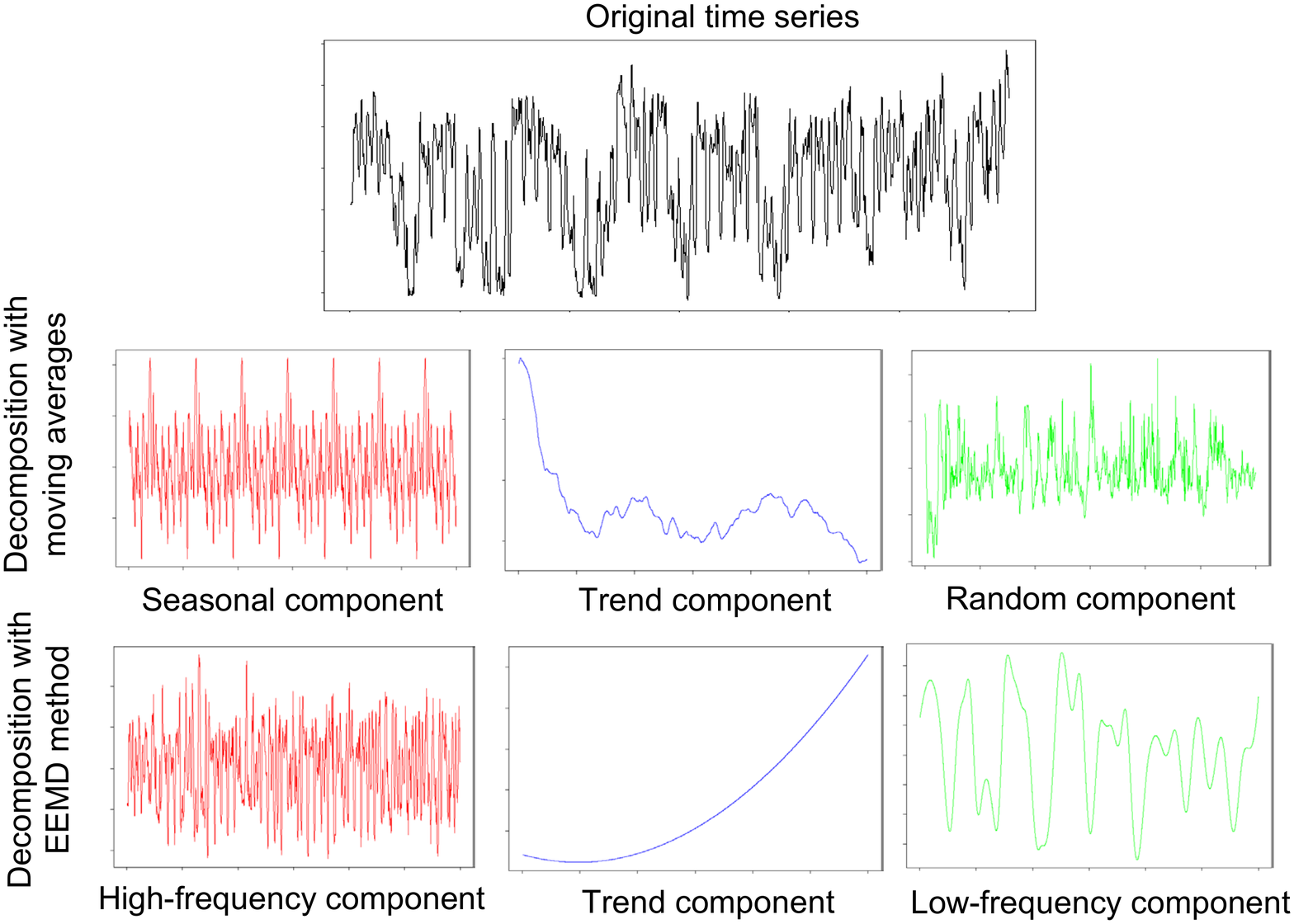}}
		\caption{Decomposed components with both approaches.}
		\label{decomp}
	\end{figure}

\begin{figure}[h]
		\centering{\includegraphics[width=.7\textwidth]{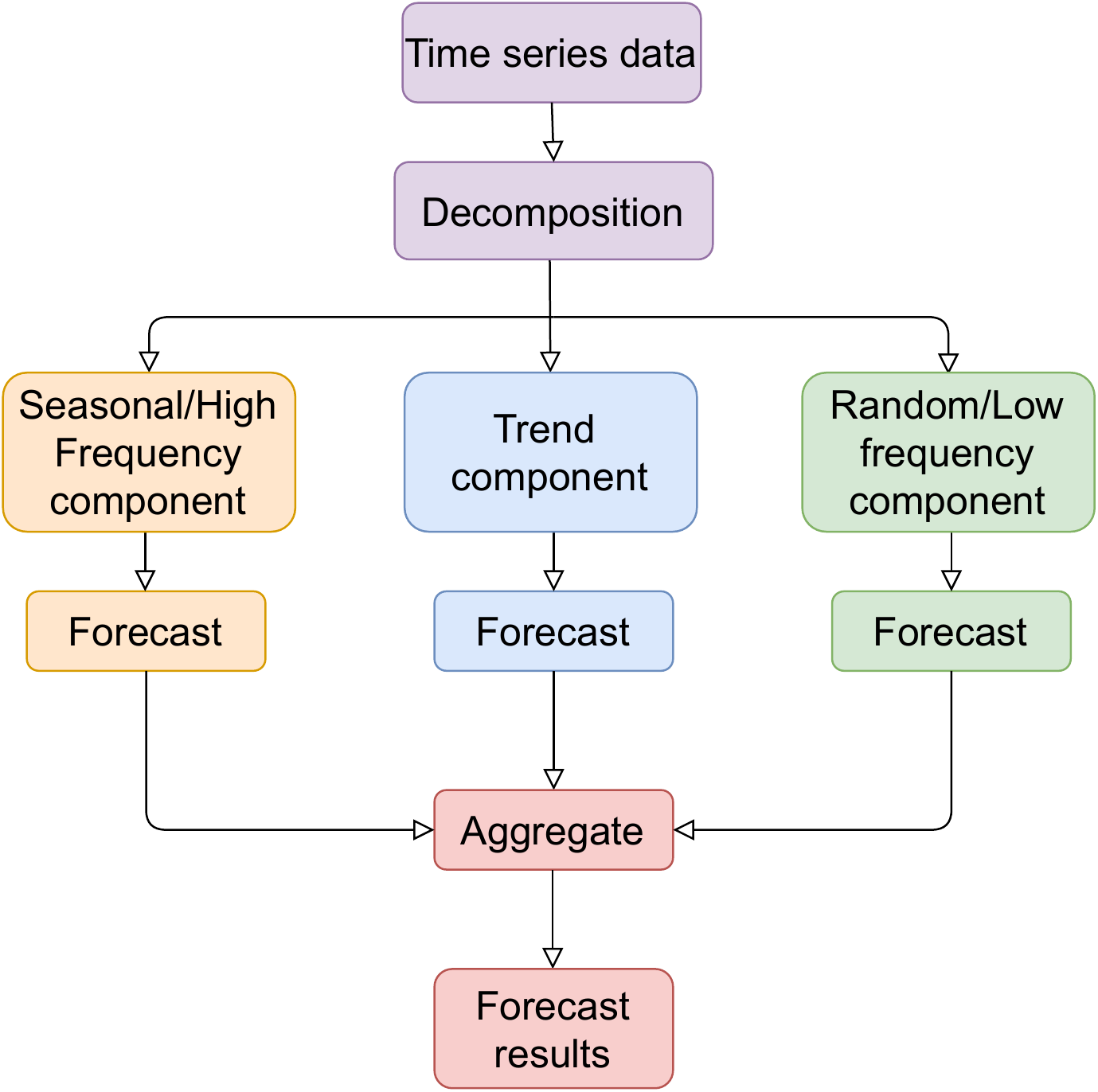}}
		\caption{Block diagram of the proposed forecasting methods.}
		\label{fig0}
	\end{figure}

	After decomposing the time series with both approaches, as shown in Figure \ref{decomp}, the decomposed component forecasting method is selected for each. The forecasted values are then aggregated to produce the final forecast results for the original time series as described in Figure \ref{fig0}. The first and second decomposition-based forecasting methods are shown in Figure \ref{fig0} (a) and (b), respectively.
	
	Now, the selection of the forecasting models for each component is a crucial process. A cross-validation approach is adopted to decide the most suitable forecasting model for each decomposed component obtained in the discussed decomposition methods. The data set used for this purpose is from the West Denmark price area (DK-1), which is the same as discussed in detail in subsequent Section 6. Ten continuous patches of time series with the length of 1200 samples are selected randomly with the Monte-Carlo strategy and decomposed into three components with both approaches as discussed in earlier sections. These components are forecasted for the next 48 values with ARIMA, FFNN, PSF, and DPSF models. Throughout the article, the comparisons of accuracies of forecasting models are executed with three error metrics; RMSE, MAE, and MAPE, which are given as:
	
	\begin{equation}
	\label{eqn_1}
	\mathrm{RMSE =  \sqrt{\frac{1}{N}\sum_{i=1}^{N}\left | X_{i}-\hat{X}_{i} \right |^{2} }}
	\end{equation}
	
	\begin{equation}
	\label{eqn_2}
	\mathrm{MAE =  \frac{1}{N}\sum_{i=1}^{N}\left | X_{i}-\hat{X}_{i} \right |}
	\end{equation}
	
	\begin{equation}
	\label{eqn_5}
	\mathrm{MAPE =  \frac{1}{N}\sum_{i=1}^{N}\frac{\left | X_{i}-\hat{X}_{i} \right |}{X_{i}} \times 100\%}
	\end{equation}
	
	\noindent where $\mathrm{X_{i}}$ and $\mathrm{\hat{X}_{i}}$ are the observed and forecasted data at time $\mathrm{t}$, respectively. $\mathrm{N}$ is the number of time steps for forecast evaluation.
	
	Table \ref{tab_e1} shows averaged RMSE and MAE values for forecasted 48 values of the time series for seasonal, trend, and random components with the first decomposition approach.
	
	\begin{table}[h]
		\caption{Averaged error values for forecasted 48 values of time series for decomposed components with moving averages with Monte-Carlo strategy for Danish dataset (* indicates best performing method. Both RMSE and MAE are values of CO2 intensity in gCO2eq/kWh).}
		\label{tab_e1}
		\makebox[\columnwidth]{
			\begin{tabular}{|c|c|c|c|c|c|c|c|c|}
				\hline
				\multirow{2}{*}{Models} & \multicolumn{2}{c}{\begin{tabular}[|c|]{@{}c@{}}Original\\ time series\end{tabular}}  & \multicolumn{2}{|c|}{\begin{tabular}[c]{@{}c@{}}Seasonal\\ component\end{tabular}} & \multicolumn{2}{|c|}{\begin{tabular}[c]{@{}c@{}}Trend\\ component\end{tabular}} & \multicolumn{2}{c|}{\begin{tabular}[c]{@{}c@{}}Random\\ component\end{tabular}} \\ \cline{2-9}
				& RMSE                                     & MAE                                     & RMSE                                    & MAE                                    & RMSE                                  & MAE                                   & RMSE                                   & MAE                                   \\ \hline
				ARIMA                   & 44.90                                    & 41.92                                   & 10.81                                   & 09.29                                  & 02.79*                                 & 02.37*                                 & 17.90*                                  & 14.70*                                 \\ \hline
				FFNN                    & 60.09                                    & 55.74                                   & 00.03*                                   & 00.02*                                  & 15.38                                 & 14.16                                 & 29.38                                  & 26.77                                 \\ \hline
				PSF                     & 43.92                                    & 36.09                                   & 01.21                                   & 00.59                                  & 04.78                                 & 03.91                                 & 21.45                                  & 18.52                                 \\ \hline
				DPSF                    & 31.88*                                    & 28.46*                                   & 09.61                                   & 08.66                                  & 03.55                                 & 03.09                                 & 19.92                                  & 17.74                                 \\ \hline
				{Mean} & \multicolumn{2}{c|}{{293.1}} & \multicolumn{2}{c|}{{7.90}} & \multicolumn{2}{c|}{{82.31}} & \multicolumn{2}{c|}{{10.88}} \\ \hline
			\end{tabular}
		}
	\end{table}
	
	From this table, it concludes that the seasonal component is efficiently predicted with the FFNN model and the other two components with distinct ARIMA models. Similarly, Table \ref{tab_e2} shows error values for forecasted 48 values with the second decomposition approach. For all cases, different ARIMA models have fitted accurately and achieved the best forecasting results. In both tables (\ref{tab_e1} and \ref{tab_e2}), the last row represents the mean of corresponding time series components.
	
	\begin{table}[h]
		\caption{Averaged error values for forecasted 48 values of time series for decomposed components with EEMD method with Monte-Carlo strategy for Danish dataset (* indicates best performing method. Both RMSE and MAE are values of CO2 intensity in gCO2eq/kWh).}
		\label{tab_e2}
		\makebox[\columnwidth]{
			\begin{tabular}{|c|c|c|c|c|c|c|c|c|}
				\hline
				\multirow{2}{*}{Models} & \multicolumn{2}{c}{\begin{tabular}[c]{@{}c@{}}Original\\ time series\end{tabular}} & \multicolumn{2}{|c}{\begin{tabular}[c]{@{}c@{}}High-frequency\\ component\end{tabular}} & \multicolumn{2}{|c}{\begin{tabular}[c]{@{}c@{}}Low-frequency\\ component\end{tabular}} & \multicolumn{2}{|c|}{\begin{tabular}[c]{@{}c@{}}Trend\\ component\end{tabular}} \\ \cline{2-9}
				& RMSE                                     & MAE                                     & RMSE                                       & MAE                                       & RMSE                                      & MAE                                       & RMSE                                  & MAE                                   \\ \hline
				ARIMA                   & 44.90                                    & 41.92                                   & 12.10*                                      & 09.07*                                     & 01.63*                                     & 01.44*                                     & 0.005*                                 & 0.003*                                 \\ \hline
				FFNN                    & 60.09                                    & 55.74                                   & 13.17                                      & 10.08                                     & 22.22                                     & 19.04                                     & 01.99                                 & 01.81                                 \\ \hline
				PSF                     & 43.92                                    & 36.09                                   & 15.59                                      & 11.32                                     & 26.28                                     & 25.93                                     & 06.33                                 & 06.32                                 \\ \hline
				DPSF                    & 31.88*                                    & 28.46*                                   & 23.01                                      & 20.80                                     & 07.93                                     & 06.39                                     & 00.05                                 & 00.04                                 \\ \hline
				{Mean} & \multicolumn{2}{c|}{{293.1}} & \multicolumn{2}{c|}{{23.11}} & \multicolumn{2}{c|}{{27.06}} & \multicolumn{2}{c|}{{10.03}} \\ \hline
			\end{tabular}
		}
	\end{table}

	\section{Performance evaluation of the proposed methods}
	
	In this section, the performance and robustness of both decomposition approaches are checked on time series forecasting methods with various distinct trends and seasonality. Then the forecasted results are compared with state-of-the-art methods.
	
	Similar to many other chaotic and intermittent time series found in nature such as wind energy, the CO$_2$ intensity time series seems to fail at achieving outstanding forecasting performance. These time series are usually infected with unusual and different patterns of trends and seasonality components. Hence, it becomes necessary to choose a method, which can handle trends and seasonality patterns individually to achieve more accurate predictions for such data.
	
\begin{figure}[h]
		\centering
		\includegraphics[width=\textwidth]{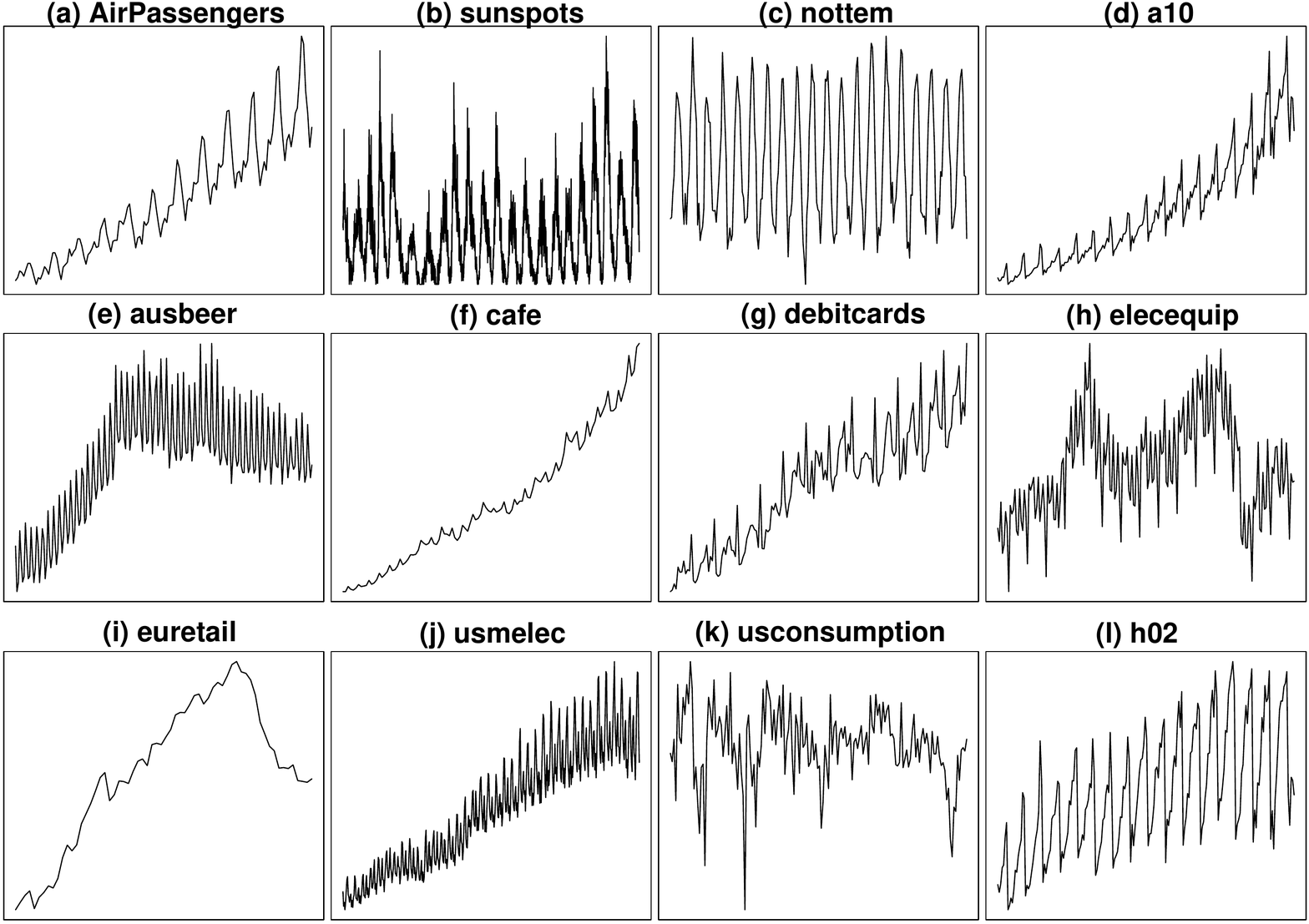}
		\caption{Time series patterns (Source: \cite{bokde2018analysis}).}
		\label{F1}
	\end{figure}

	Before the CO$_2$ intensity data set, the performance of proposed methods based on two mentioned decomposition methods is evaluated for some standard time-series data sets, which are available under a public license. The data sets, used here, are listed in Table \ref{A1} and Figure \ref{F1}. Additionally, the statistical characteristics of the respective time series are provided in Table \ref{A2n}. This table provides the basic statistical details of time series that includes minimum (Min.) and maximum (Max.) values, Median, Mean, 1st and 3rd Quartiles.
	
	\begin{table}[h]
		\footnotesize
		\centering
		\caption{Details of time series used for validation (Source: \cite{bokde2018analysis}).}
		\label{A1}
		\makebox[\columnwidth]{
			\begin{tabular}{|c|l|}
				\hline
				\multicolumn{1}{|c|}{\textbf{\begin{tabular}[c]{@{}c@{}}Time Series\end{tabular}}} & \multicolumn{1}{c|}{\textbf{Description}}                                                                                                                                               \\ \hline
				\textbf{AirPassangers}                                                                         & Time series for monthly totals of international airline passengers for 11 years  \\ \hline
				\textbf{nottem}                                                                         & \begin{tabular}[c]{@{}l@{}}Time series for average air temperature at Nottingham Castle in \\ degree Fahrenheit for 20 years.\end{tabular} \\ \hline
				\textbf{sunspots}                                                                         & Time series for monthly mean relative sunspot numbers for more than two centuries. \\ \hline
				\textbf{a10}                                                                         & \begin{tabular}[c]{@{}l@{}}Total monthly scripts for pharmaceutical products falling under ATC code A10. \\ It is collected by the Australian Health Insurance Commission.\end{tabular} \\ \hline
				\textbf{ausbeer}                                                                     & \begin{tabular}[c]{@{}l@{}}Total quarterly beer production in Australia (in megalitres) reported for more \\ than 50 years.\end{tabular}                                                \\ \hline
				\textbf{cafe}                                                                        & \begin{tabular}[c]{@{}l@{}}Total quarterly expenditure on cafes, restaurants and takeaway food services \\ in Australia collected for around 30 years.\end{tabular}                     \\ \hline
				\textbf{debitcards}                                                                  & Retail debit card usage in Iceland (million ISK).                                                                                                                                       \\ \hline
				\textbf{elecequip}                                                                   & \begin{tabular}[c]{@{}l@{}}Manufacture of electrical equipment: computer, electronic and optical \\ products.\end{tabular}                                                              \\ \hline
				\textbf{euretail}                                                                    & \begin{tabular}[c]{@{}l@{}}Quarterly retail trade index in the Euro area in 17 countries, from 1996 \\ to 2011.\end{tabular}                                                            \\ \hline
				\textbf{h02}                                                                         & \begin{tabular}[c]{@{}l@{}}Total monthly scripts for pharmaceutical products falling under ATC code H02, \\ as recorded by the Australian Health Insurance Commission.\end{tabular}     \\ \hline
				\textbf{usconsumptions}                                                              & \begin{tabular}[c]{@{}l@{}}Percentage changes in quarterly personal consumption expenditure and personal \\ disposable income for the US, 1970 to 2010.\end{tabular}                    \\ \hline
				\textbf{usmelec}                                                                     & Electricity net generation measured in billions of kilowatt hours (kWh).                                                                                                                \\ \hline
			\end{tabular}
		}
	\end{table}
	
	\begin{table}[h]
		\footnotesize
		\centering
		\caption{Statistical characteristics of time series used for validation (Source: \cite{bokde2018analysis}).}
		\label{A2n} 
			\begin{tabular}{|l|c|c|c|c|c|c|}
				\hline
				\multicolumn{1}{|c|}{\textbf{Time Series}} & \textbf{Min.} & \textbf{1st Qu.} & \textbf{Median} & \textbf{Mean}& \textbf{3rd Qu.} & \textbf{Max.} \\ \hline
				\textbf{AirPassengers}                     &104.0  & 180.0  & 265.5 &  280.3 &  360.5 &  622.0        \\ \hline
				\textbf{sunspots}                          &0.00  & 15.70 &  42.00 &  51.27 &  74.92 & 253.80       \\ \hline
				\textbf{nottem}                            & 31.30 &  41.55 &  47.35 &  49.04 &  57.00 &  66.50 \\ \hline
				\textbf{a10}                               & 2.815 &  5.844 &  9.319 & 10.694 & 14.290 & 29.665  \\ \hline
				\textbf{ausbeer}                           & 213.0 &  378.5 &  423.0 &  415.0 &  465.5 &  599.0  \\ \hline
				\textbf{cafe}                              & 1012  &  2190  &  3389  &  3716  &  5193  &  8426\\ \hline
				\textbf{debitcards}                        & 7204  & 11330  & 16183  & 15514  & 18514  & 26675\\ \hline
				\textbf{elecequip}                         & 62.47 &  86.17 &  94.03 &  95.71 & 103.95 & 128.75\\ \hline
				\textbf{euretail}                          & 89.13 &  94.56 &  96.55 &  96.35 &  99.43 & 102.10\\ \hline
				\textbf{h02}                               & 0.3362 &  0.5889 & 0.7471 & 0.7682 & 0.9376 & 1.2572\\ \hline
				\textbf{usconsumptions}                    & -2.2966 & 0.4050 & 0.8005 & 0.7553 & 1.1141 & 2.3171\\ \hline
				\textbf{usmelec}                           &139.6  & 192.6 &  252.5 &  254.1 &  307.0 &  421.8  \\ \hline
		\end{tabular} 
	\end{table}
	
	These {time-series} data sets were used for performance evaluation of a forecasting model in \cite{bokde2018analysis}. The selection of these time series is crucial because special precautions were taken to consider various kinds of seasonality, trend components as well as a variety of periodic and chaotic data sets, which are further discussed in \cite{bokde2018analysis}. The amount of chaos in the time series data sets is quantified in terms of the maximum Lyapunov exponent (max. LE) values. The max. LE value of the `sunspots’ time series is 0.6009, which makes it a highly chaotic series, whereas, the `nottem’ time series is highly periodic with max. LE value of 0.0065. For all of these time series, the last 48 values are taken away for validation purpose, whereas the leftover initial part of the {time-series is} used for training of the forecasting methods. Then the proposed methods are compared with ARIMA, DPSF, FFNN, and PSF methods. The R package `forecast’ \cite{hyndman2007automatic} is used for ARIMA and FFNN; and `PSF’ \cite{bokde4psf, bokde2017psf} is used for PSF methods.
	
	Table \ref{A2} shows a comparison of the proposed decomposition methods in terms of RMSE values with other methods in the study of a 48-hour forecast. It is observed that the proposed methods achieved better results for almost all {time-series}. Further, the accuracy of the proposed methods is backed by Friedman's non-parametric test with the distribution of predicted observations. This test is generally used to find differences in treatments within multiple outcomes. Non-parametric means the test does not assume that data belongs to a particular distribution. In this test, the null hypothesis assumes that data comes from a particular distribution. The null hypothesis of this test states that samples in two or more groups are drawn from the same mean value populations. Further, the null hypothesis is rejected when the level of significance is below 0.05.
	
	\begin{table}[h]
		\footnotesize
		\centering
		\caption{Comparison of proposed methods in terms of RMSE values (* indicates best performing method).}
		\label{A2}
		\makebox[\columnwidth]{
			\begin{tabular}{|l|c|c|c|c|c|c|}
				\hline
				\multicolumn{1}{|c|}{\textbf{Time Series}} & \textbf{ARIMA} & \textbf{FFNN} & \textbf{PSF} & \textbf{DPSF}& \textbf{Method 1} & \textbf{Method 2} \\ \hline
				\textbf{AirPassengers}                     & 113.72          & 78.72             & 97.80        & 83.09 & 42.19* & 65.31        \\ \hline
				\textbf{sunspots}                          & 103.07          & 32.24             & 52.29        & 42.51* & 47.03 & 59.34         \\ \hline
				\textbf{nottem}                            & 2.52           & 2.46              & 2.13         & 2.18 & 2.01* & 2.87         \\ \hline
				\textbf{a10}                               & 4.88           & 4.20              & 6.66         & 4.65 & 2.83* & 3.18         \\ \hline
				\textbf{ausbeer}                           & 33.23          & 33.19              & 76.85        & 85.95 & 30.18* & 31.12        \\ \hline
				\textbf{cafe}                              & 1208.62         & 2534.18            & 2888.15      & 1208.62 & 1195.02 & 380.44*        \\ \hline
				\textbf{debitcards}                        & 2638.13        & 3353.65            & 3060.86      & 6664.44 & 1029.00* & 3069.27       \\ \hline
				\textbf{elecequip}                         & 27.76           & 15.80              & 15.57         & 32.27 & 20.79 & 10.82*         \\ \hline
				\textbf{euretail}                          & 06.17           & 04.52              & 12.39         & 08.32   & 03.13* & 05.29       \\ \hline
				\textbf{h02}                               & 0.22           & 0.17*              & 0.08         & 0.18 & 0.17* & 0.18          \\ \hline
				\textbf{usconsumptions}                    & 0.98           & 0.96*              & 1.15         & 1.10 & 0.99 & 0.99          \\ \hline
				\textbf{usmelec}                           & 38.33          & 15.12*             & 50.22        & 34.17 & 21.71 & 34.27         \\ \hline
			\end{tabular}
		}
	\end{table}
	
	The p-values for all-time series for Friedman’s test for the proposed methods are shown in Table \ref{friedman}. In all cases, the null hypothesis is rejected with p-values lower than 0.05. This concludes that the proposed methods have achieved statistically significant outcomes.
	
	\begin{table}[h]
		\centering
		\caption{p-values of Friedman's test for proposed methods.}
		\label{friedman}
		\begin{tabular}{|l|c|c|c|l|l|}
			\hline
			\multicolumn{1}{|c|}{\textbf{Time Series}} & \textbf{Method 1} & \textbf{Method 2} \\ \hline
			\textbf{AirPassengers}                     & 1.749e-04	& 2.871e-12             \\ \hline
			\textbf{sunspots}                          & 7.764e-09  & 4.262e-12                   \\ \hline
			\textbf{nottem}                            & 2.400e-03		& 1.222e-12        \\ \hline
			\textbf{a10}                               & 1.490e-05   & 1.902e-12               \\ \hline
			\textbf{ausbeer}                           & 2.141e-10  & 3.551e-12           \\ \hline
			\textbf{cafe}                              & 7.764e-09	& 1.401e-03             \\ \hline
			\textbf{debitcards}                        & 4.331e-03	& 4.262e-12               \\ \hline
			\textbf{elecequip}                         & 7.764e-09	& 4.002e-12                \\ \hline
			\textbf{euretail}                          & 3.147e-11	& 4.139e-08                  \\ \hline
			\textbf{h02}                               & 4.139e-08	& 1.974e-12             \\ \hline
			\textbf{usconsumptions}                    & 1.400e-03 	& 3.892e-03                  \\ \hline
			\textbf{usmelec}                           & 5.312e-05	& 4.262e-12                \\ \hline
		\end{tabular}
	\end{table}
	
	\section{CO$_2$ intensity forecast. A case study}
	In this section, the performance of the proposed methods is examined for short-term CO$_2$ intensity forecasting with the consideration of two decomposition approaches. As discussed in earlier sections, these methods are based on two decomposition approaches which decompose a time series into three distinct components. The comparison of these methods is applied for CO$_2$ intensity forecasting for 48 hours ahead with recursive and Dirrec strategies of predictions. In a recursive approach, the first value is predicted with a prediction model and then appended to the data set. The same model is then used to predict the next value. In the Dirrec approach, the first value is predicted with a model and appended to the data set. A new model is then fitted and used to predict the next value and so on. As the Dirrec approach is a more time-consuming technique, it will be interesting to understand its effect on forecasting accuracy.
	
	\subsection{Data}
	The data set used in this study is provided by the ENSTO-E Transparency platform \cite{transparency}. Transmission system operators of each price area report data for production mix, demand, import and export flows{, and} prices to this platform on daily intervals with hourly resolution. A visualization of this data is provided by the electricityMap \cite{electricitymap}. This data for 2017 was the foundation for \cite{tranberg2019real}, where a real-time carbon accounting method was proposed for the European electricity markets. This method was based on power flow tracing techniques and introduced a new consumption-based accounting method that represents the underlying physics of the electricity system physics in contrast to the state-of-the-art input-output carbon accounting models \cite{fan2016exploring, zhang2018production, clauss2018generic}.
	
	In this case study, {data is used from} the ENSTO-E Transparency platform for 2018 and 2019. The data set includes electricity generation per technology and electricity demand per price area{, as well as} power flows between interconnected areas. {The flow tracing is used} to map power flows between importing and exporting countries as described in \cite{Tranberg2015}. Country-specific average CO$_2$ emission intensity per generation technology is applied to the flow tracing results to calculate the hourly CO$_2$ intensity of electricity consumption for each price area following the method proposed in \cite{tranberg2019real}. {It} refer to this as the \textit{consumption intensity}. Based on the local generation mix, the CO$_2$ intensity of electricity generation for each price area is referred to as the \textit{production intensity}.
	
\begin{figure}[p]
		\centering{\includegraphics[width=\textwidth]{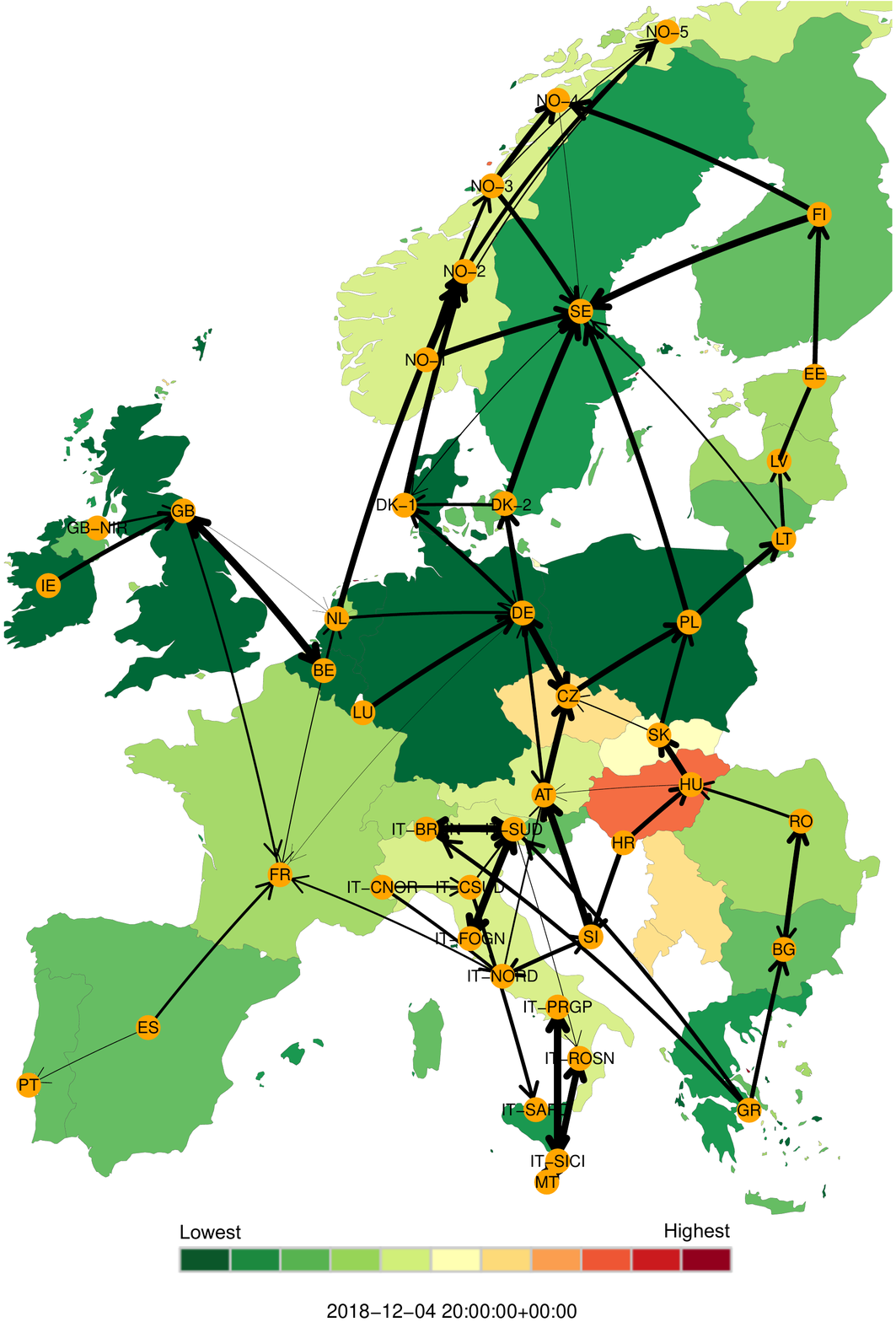}}
		\caption{The countries with carbon consumption intensities (shown in different color ranges) and their inter-connectors (indicating import and export of power flow) for a specific hour.}
		\label{fig6}
	\end{figure}

	Figure \ref{fig6} shows the countries and their inter-connectors (indicating import and export of power flow) for a specific hour considered in the present study. The width of inter-connectors indicates the proportions of the power flow within the countries. And the color allotted to each country represents the consumption-based CO$_2$ intensity for the specific country within the given color palette. The color sheds vary from green to red color indicating low to high CO$_2$ intensities for the specific hour.
	
\begin{figure}[h]
		\centering{\includegraphics[width=\textwidth]{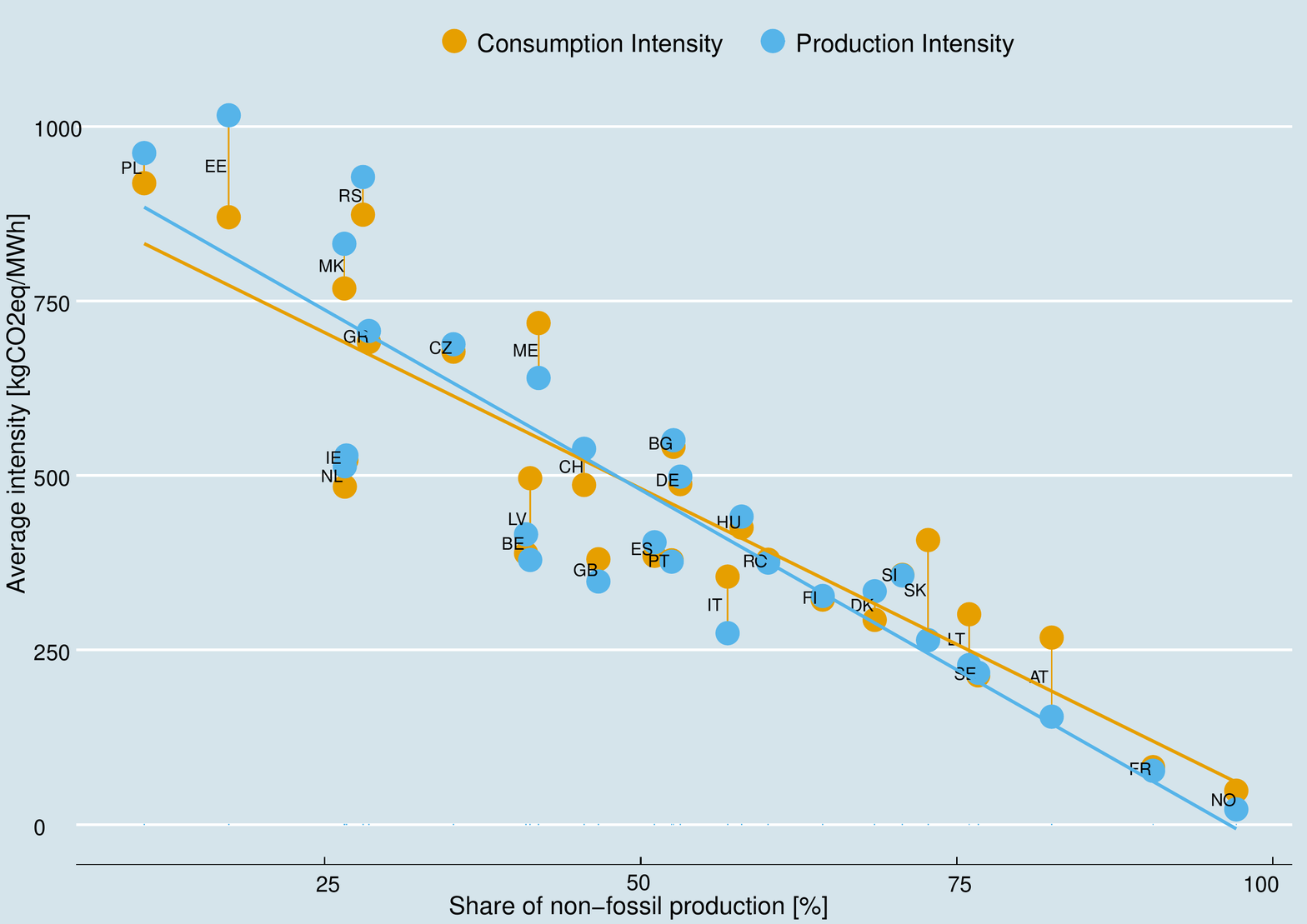}}
		\caption{For each country, {the compared} average hourly production (blue) and consumption (orange) intensity as a function of the share of non-fossil generation in the country's generation mix.}
		\label{fig6x}
	\end{figure}

	\newpage
	In Figure \ref{fig6x}, the average hourly production and consumption intensity {are compared} for each country as a function of the share of non-fossil generation in the country's generation mix. A clear pattern of decreasing intensity {is seen} with {an increasing} share of non-fossil generation. With few exceptions, the consumption intensity is higher than the production intensity for high shares of non-fossil generation. This is an indication of the imported electricity having a higher intensity than the local generation on average. The pattern reverses for low shares of non-fossil generation.

\begin{figure}[h]
		\centering{\includegraphics[width=\textwidth]{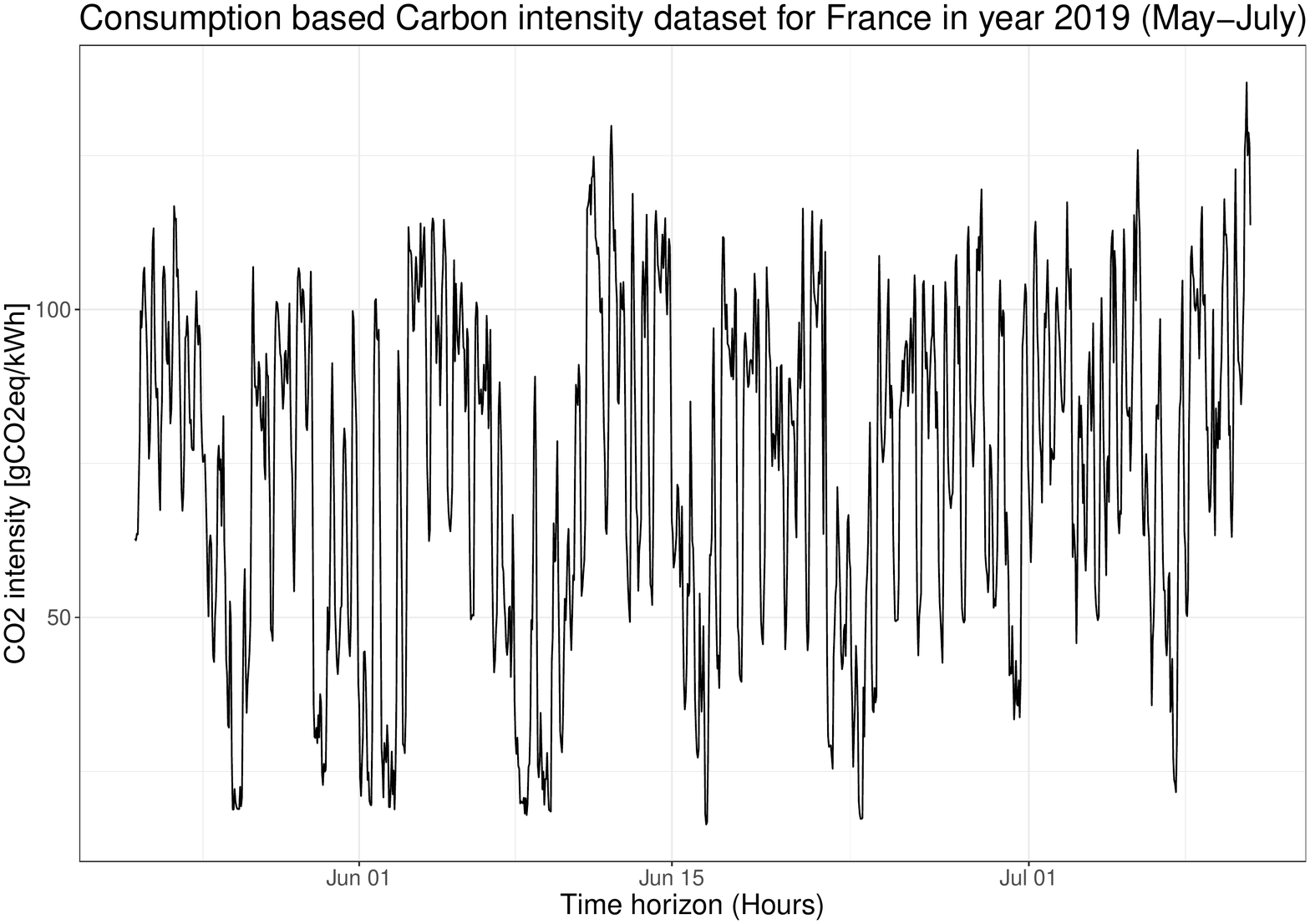}}
		\caption{CO$_2$ intensity time series data set from France for 1248 hours (52 days).}
		\label{francets}
	\end{figure}

	For the detailed demonstration, CO$_2$ intensity data from 25-05-2019 to 10-07-2019 (i.e. 1248 hours = 52 days) for France is selected randomly for analysis as illustrated in Figure \ref{francets}. This time series is divided into two parts, i.e., training data which contains data of initial 1200 sampling (1200 hours = 50 days) and validation data as the last 48 samplings (48 hours = 2 days). Further, Table \ref{spd_1} shows the statistical characteristics of the training data set.
	
	\begin{table}[h]
		\footnotesize
		\centering
		\caption{Statistical characteristics of CO$_2$ intensity time series data set from France.}
		\label{spd_1}
		\begin{tabular}{|c|c|c|c|c|c|}
			\hline
			\textbf{Location} & \textbf{Mean} & \textbf{Median} & \textbf{Minimum} & \textbf{Maximum} & \textbf{Standard Deviation} \\ \hline
			France  & 74.16        & 78.93           & 16.35             & 136.89           & 27.01                     \\ \hline
		\end{tabular}
	\end{table}

	\subsection{Model selection}
	Forecast result accuracy is primarily dependent on the proper selection of models. In this section, {the} model selection of the proposed methods and state-of-the-art methods are discussed.
	These state-of-the-art methods are modeled for the original time series as follows.
	ARIMA models are consist of three parameters, $p$ (number of auto-regressive terms), $d$ (number of nonseasonal differences needed for stationarity) and $q$ (number of lagged forecast errors in the prediction equation); for the given time series these values are 2, 1 and 1, respectively. For PSF models, the optimum window ($W$) and cluster size ($K$) parameters are derived with the R package PSF \cite{bokde2017psf} and for given time series, these values are 4 and 3, respectively. Similarly, for the DPSF model, these values are 5 and 3, respectively. For the FFNN model, two parameters ($p$ lagged inputs and $k$ nodes in the single hidden layer) are calculated and for a given time series, these values are 26 and 14, respectively.
	
\begin{figure}
		\centering
		\begin{subfigure}[b]{\textwidth}
			\centering{\includegraphics[width=\textwidth]{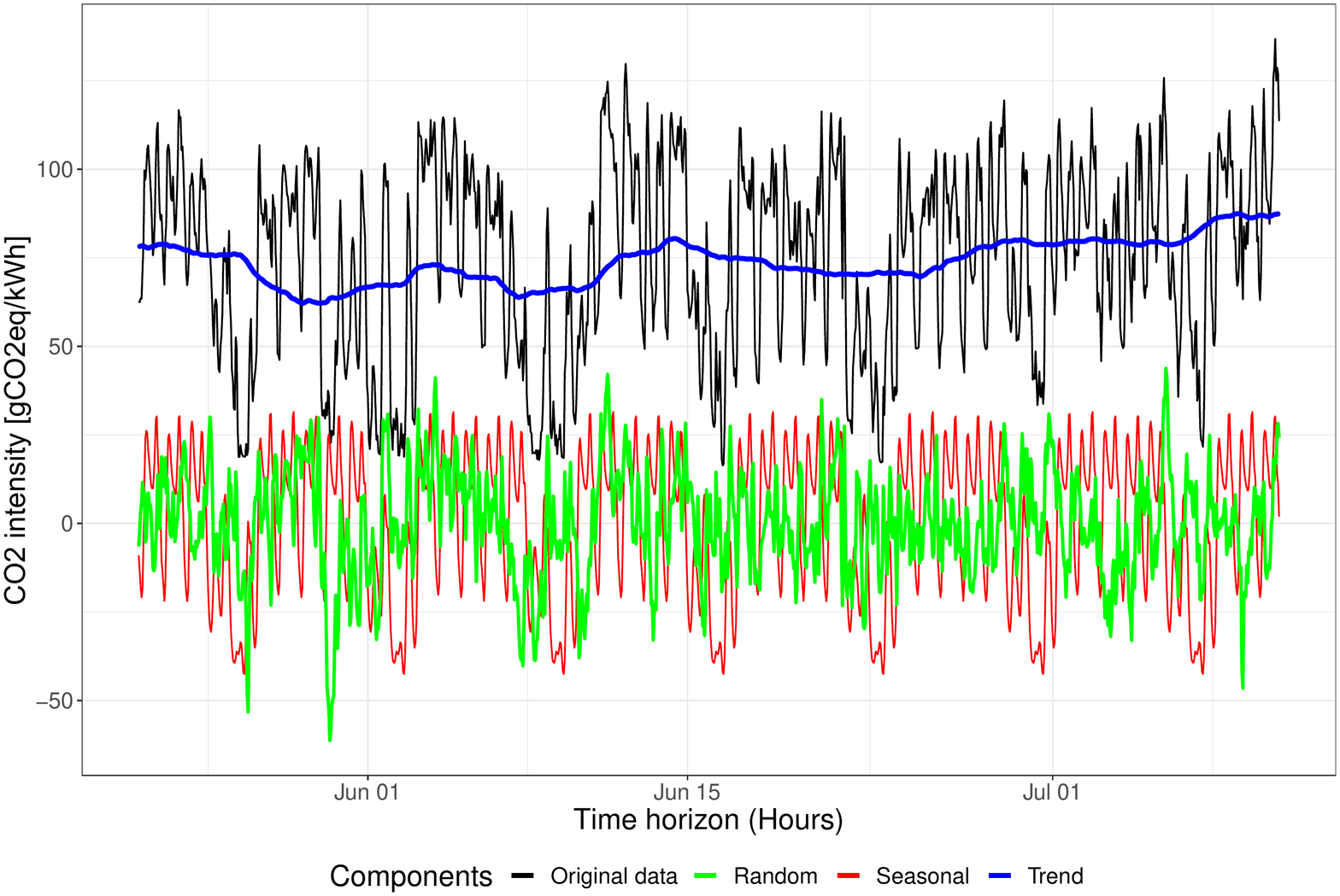}}
			\caption{Decomposition with the moving averages.}
			\label{Fra_M1}
		\end{subfigure}
		\begin{subfigure}[b]{\textwidth}
			\centering{\includegraphics[width=\textwidth]{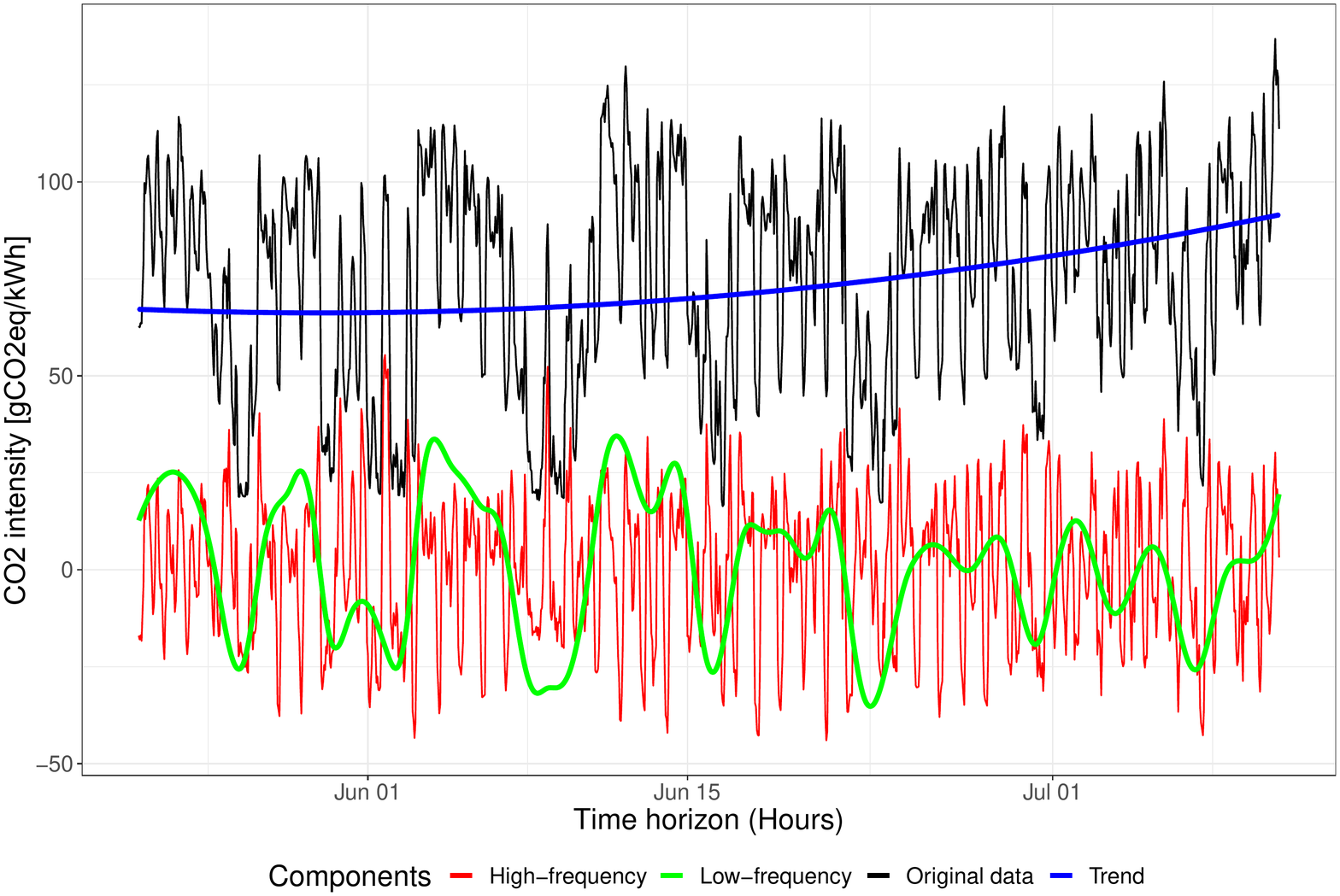}}
			\caption{Decomposition with EEMD method.}
			\label{Fra_M2}
		\end{subfigure}
		\caption{Decomposed components obtained with the (a) moving averages and (b) EEMD method based decomposition approaches.}
		\label{fig:three graphs}
	\end{figure}

	Further, for the first proposed method, there are three models for three components (i.e. seasonal, trend{, and} random components) and are shown in Figure \ref{Fra_M1}. For the seasonal component, FFNN is modeled with $p=28$ and $k=14$. Whereas, trend and random components are forecasted with ARIMA models. The corresponding parameters are $p =1$, $d =1$, $q =0$ and $p =3$, $d =0$, $q =1$, respectively.

\begin{figure}[h]
		\centering{\includegraphics[width=\textwidth]{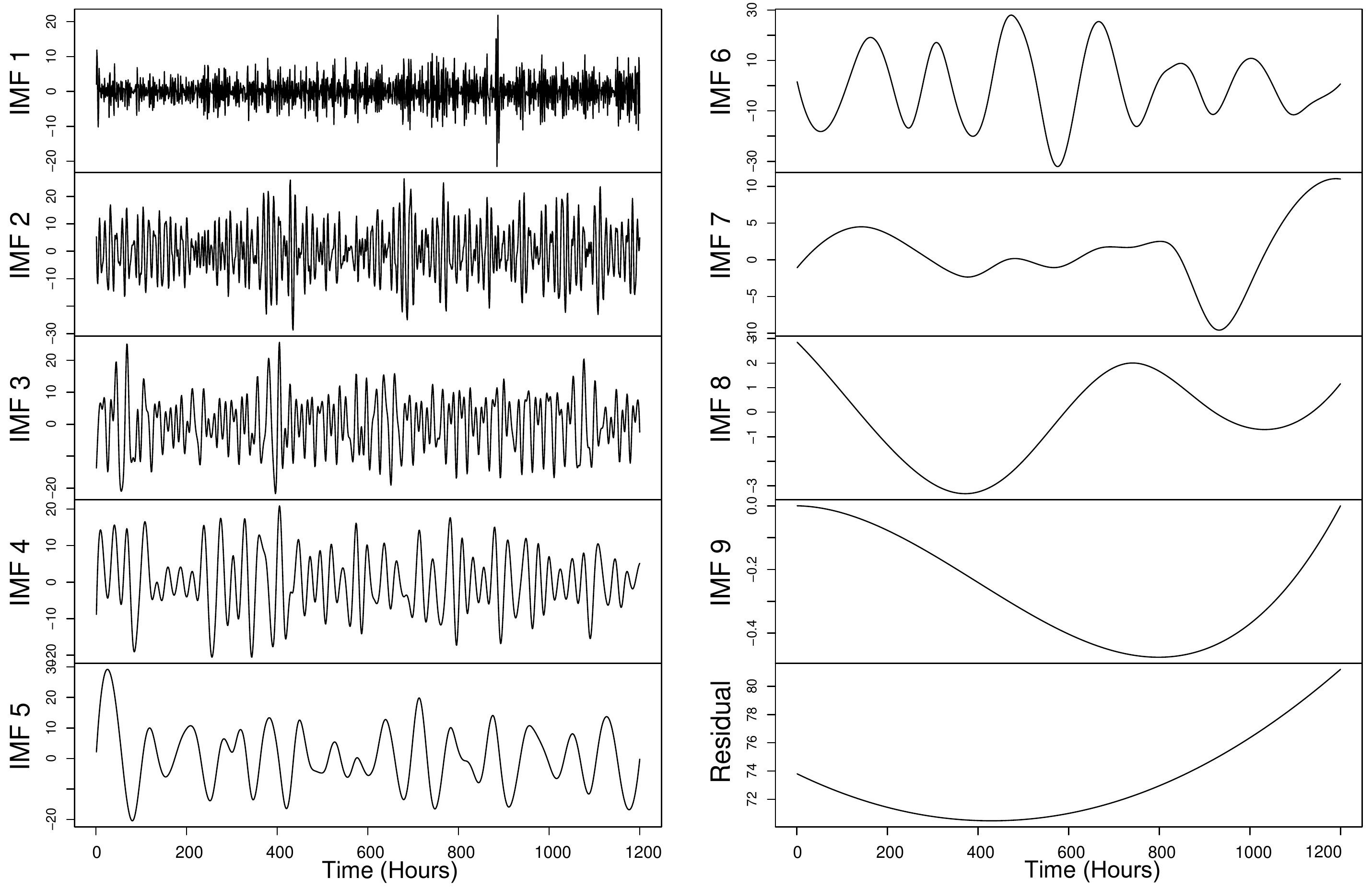}}
		\caption{Decomposition of {CO$_2$ intensity [gCO2eq/kWh]} time series for France.} 
		\label{eemd_fra}
	\end{figure}

\begin{figure}[h]
		\centering{\includegraphics[width=\textwidth]{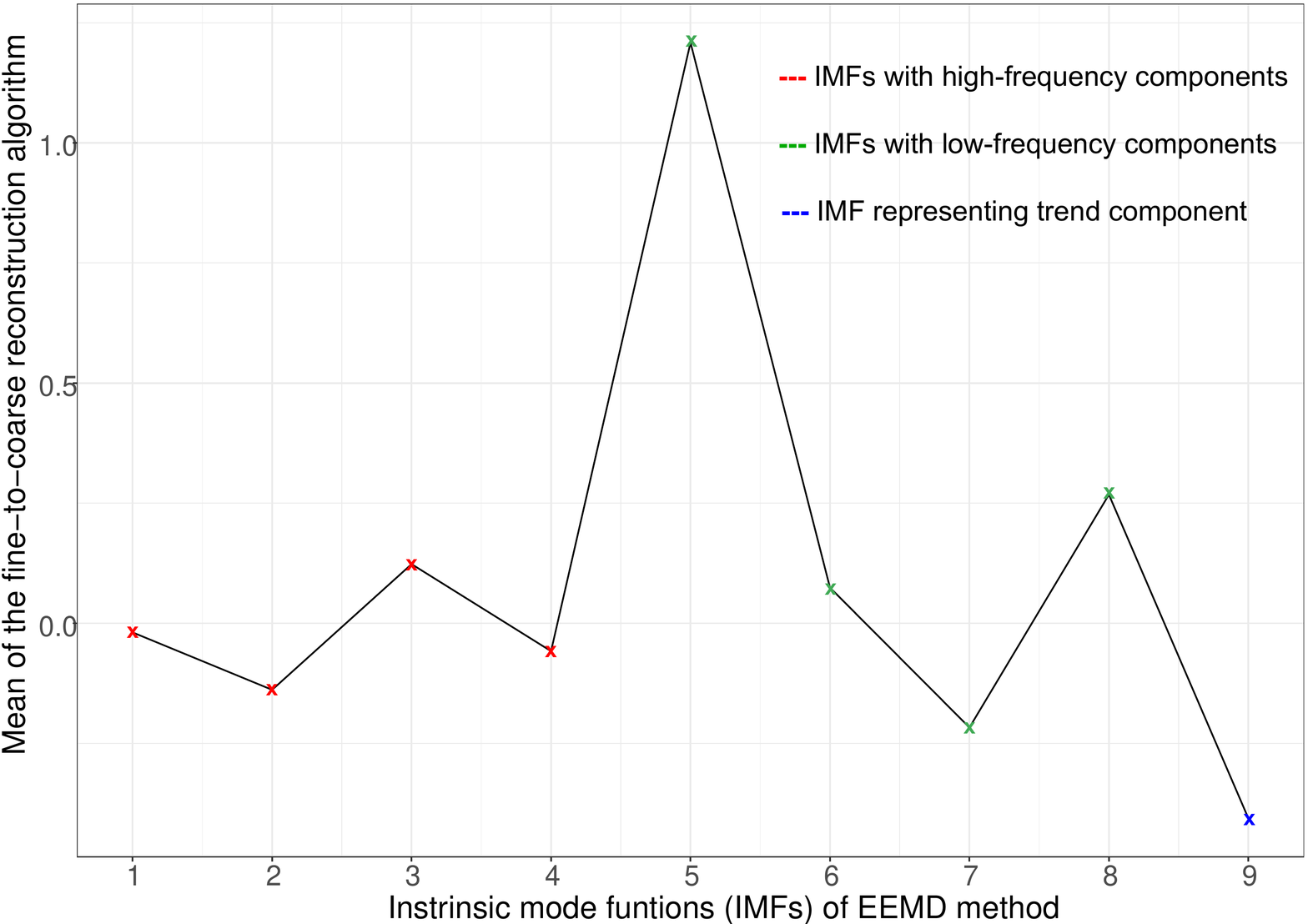}}
		\caption{The mean of the fine-to-coarse reconstruction algorithm as a function of index $K$.}
		\label{fine_to_c}
	\end{figure}

	Similarly, in the second proposed method, the time series first decomposes into the number of sub-series with EEMD, named as IMFs. For the given time series, the 10 generated IMFs are shown in Figure \ref{eemd_fra}. Then, with the help of the fine-to-course reconstruction method, all IMFs are combined into three components based on the mean values of IMFs shown in Figure \ref{fine_to_c}. Then all three components (i.e. low-frequency, high-frequency{, and} trend components) shown in Figure \ref{Fra_M2} are forecasted with ARIMA models with parameters as shown in Table \ref{hlr}.

	\begin{table}[h]
		\centering
		\caption{Parameters for ARIMA models in Method 2.}
		\label{hlr}
		\begin{tabular}{|c|c|c|c|c|}
			\hline
			Components     & Models & p & d & q \\ \hline
			High-frequency & ARIMA  & 2 & 0 & 3 \\ \hline
			Low-frequency  & ARIMA  & 1 & 0 & 0 \\ \hline
			Trend          & ARIMA  & 0 & 2 & 0 \\ \hline
		\end{tabular}
	\end{table}
	
	\subsection{Results and discussion}
	
	This subsection discusses the forecasting results of both proposed methods on the CO$_2$ intensity time series data set from several European countries. Apart from the comparison of both {decomposition-based} methods, other state-of-the-art forecasting methods have been compared in the study to prove the {accuracies} of both strategies. These models include ARIMA, FFNN, SVM, PSF, and DPSF, which are among the best ones to perform on such chaotic and intermittent data set used in this study.
	All these models are compared with both recursive and Dirrec prediction approaches for 48 hours ahead forecasting results for the French data set and are shown in Table \ref{tr1}. For both approaches, the proposed models show better forecasting performance in terms of RMSE, MAE, and MAPE with Method 1 outperforming Method 2. There {are} several models to forecast one or two-step ahead values precisely.

	\begin{table}[]
		\centering
		\caption{Comparison of proposed methods with state-of-the-art models for CO$_2$ intensity data in France (* indicates the best performing method).}
		\label{tr1}
		\begin{tabular}{|c|c|c|c|c|c|c|}
			\hline
			\multirow{2}{*}{Models} & \multicolumn{3}{c|}{Recursive approach} & \multicolumn{3}{c|}{Dirrec approach} \\ \cline{2-7}
			& RMSE        & MAE         & MAPE       & RMSE       & MAE        & MAPE      \\ \hline
			ARIMA                   & 28.86       & 22.83       & 21.64      & 28.34      & 22.41      & 21.40     \\ \hline
			FFNN                    & 26.43       & 20.28       & 21.14      & 25.93      & 19.73      & 20.88     \\ \hline
			SVM                     & 62.96       & 50.50       & 59.20      & 58.11      & 43.90      & 55.05     \\ \hline
			PSF                     & 20.47       & 15.31       & 15.26      & 20.38      & 15.29      & 15.21     \\ \hline
			DPSF                    & 26.12       & 21.93       & 22.62      & 26.05      & 21.88      & 22.56     \\ \hline
			Method 1                & 15.57*       & 11.99*       & 11.47*      & 15.53*      & 11.80*      & 11.46*     \\ \hline
			Method 2                & 18.28       & 14.58       & 15.98      & 18.16      & 14.43      & 15.91     \\ \hline
		\end{tabular}
	\end{table}

	%
	%
	
\begin{figure}[h]
		\centering{\includegraphics[width=\textwidth]{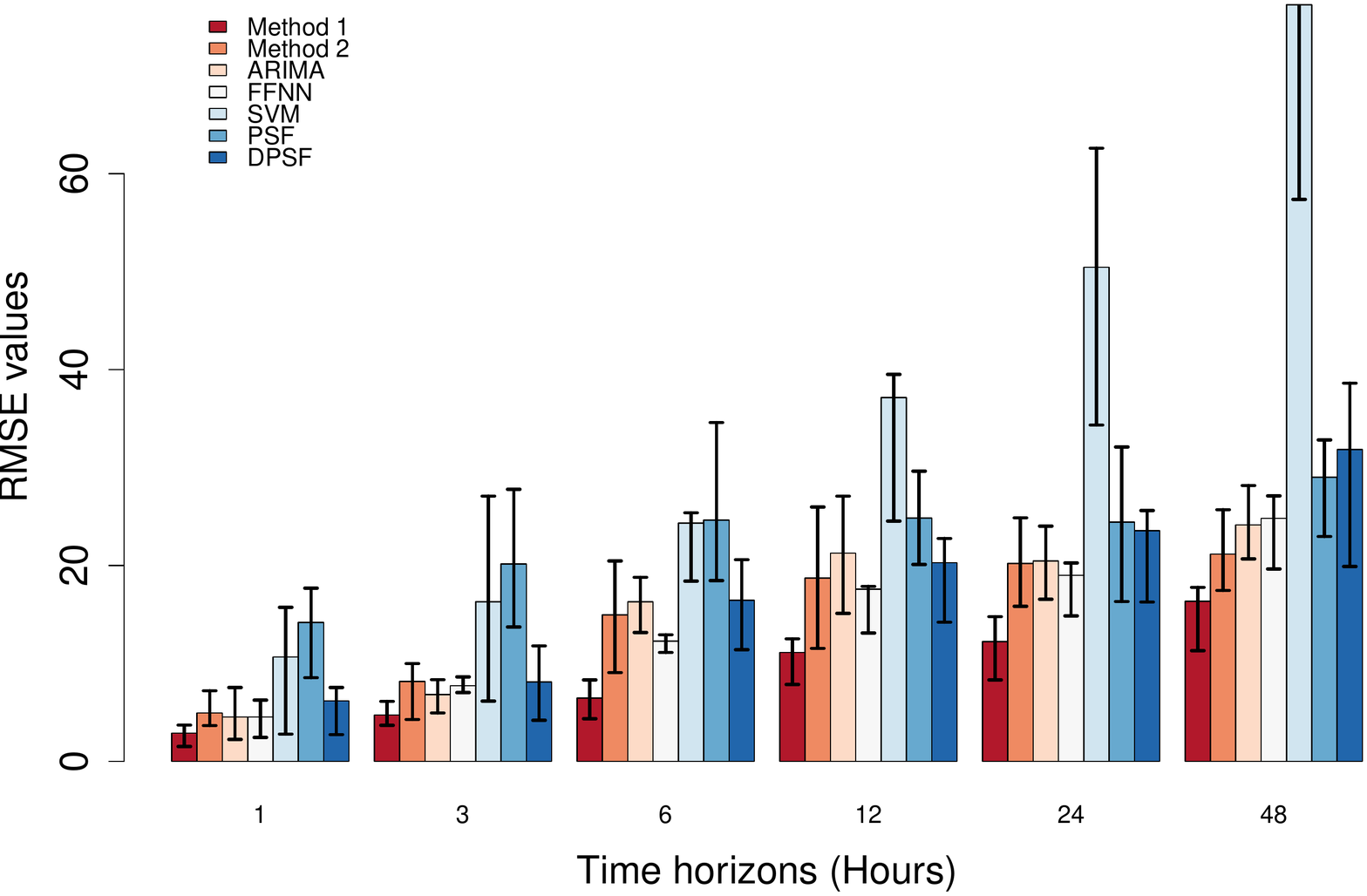}}
		\caption{Multi-step ahead error values (RMSE) with various models for France.}
		\label{Fra_horz}
	\end{figure}

	Furthermore, all these models are examined for the same French data set for different multi-step ahead forecasting. Figure \ref{Fra_horz} shows the error values (RMSE) with distinct models for 1, 3, 6, 12, 24 and 48 hours ahead forecasts. It can be seen that concerning the proposed methods, ARIMA, PSF, and DPSF models are showing lower or comparable errors for shorter multi-step forecasts, but it gets drastically higher for longer steps (48-step ahead) ones. Whereas, for 48 hours ahead forecast, both proposed models have shown higher accuracy, and especially Method 1.
	
	Apart from this, the statistical significance of the proposed methods is evaluated with Friedman’s test. The p-values of this test for the French data set are 0.020 and 0.057 for the 48 hours ahead forecast for Method 1 and Method 2, respectively. In both cases, the null hypothesis is rejected with lower p-values and approves its statistical significance.
	
	In such studies, the complexity of a methodology is considered as an important metric apart from forecasting error comparison to evaluate the forecasting performance. Usually, the complexity is estimated only with the size of the data set, not other possible parameters such as the number of features, tuning parameters, etc. In the present study, the time complexity of the proposed methods is estimated with the GuessCompx tool \cite{GuessCompx}. This tool empirically guesses the time and memory complexities of a methodology. It tests multiple, increasing size random samples of the input data sets and attempts to fit distinct complexity function $O(N)$, $O(N^2)$, $O(log N)$, etc. Based on the best fit, this tool estimates the full computational time for given data set and it is discussed in detail in \cite{agenis2019guesscompx}.
	
	Table \ref{tr3} shows the time and memory complexities of individual decomposition approaches and forecasting models for each decomposed component for both proposed methods for the French data set. The overall complexities of both methods are calculated by observing the complexities of each sub-step involved in the methods. This individual complexity study is done since all sub-steps (decomposition and forecasting models) are performed on different data series.
	
	The overall complexity of the methods is finalized based on the most complex process within each method. In Method 1, the seasonal component forecasting with the FFNN model is the most complex step with quadratic ($O(N^2)$) nature in terms of time complexity. Hence, the overall time complexity of Method 1 is quadratic ($O(N^2)$). Similarly, the decomposition approach in Method 2 appeared to be the most time consumption-based complex task, which is Linearithmic $O(N log N)$. Furthermore, the memory complexity is observed to be the Logarithmic $O(log N)$ for both methods. Considering the same length of time series, similar complexity observations are obtained for the other countries.
	
	An unbiased comparison study is performed with the Monte-Carlo strategy for all countries’, in which 25 different patches of 1248 continuous samples are selected randomly. For each patch, initial 1200 samples are used as training data and the remaining 48 values are used for validation purposes. Finally, the mean of error values for each patch is considered as the performance of the methods compared in the study. This Monte-Carlo strategy is implemented with the ForecastTB tool \cite{bokde2020forecasttb, FTB}, and it ensures the suitability of the models for the given time series and rejects the possibility of favorable results by chance.

	\begin{table}[]
		\centering
		\caption{Comparison of proposed methods with state-of-the-art models in terms of RMSE for CO$_2$ emissions data for European countries (* indicates the best performing method for the respective country).}
		\label{tr2}
		\makebox[\columnwidth]{
			\begin{tabular}{|c|c|c|c|c|c|c|c|}
				\hline
				Country          & ARIMA  & FFNN  &SVM & PSF    & DPSF   & Method 1 & Method 2 \\ \hline
				Austria         & 52.42  & 50.93  &51.66 & 57.20  & 51.82  & 35.49*    & 46.34    \\ \hline
				Belgium         & 191.75 & 200.17 &298.35 & 223.17 & 321.50 & 195.17   & 181.22*   \\ \hline
				Bulgaria        & 36.40  & 55.15  &85.12 & 61.21  & 43.14  & 32.98    & 31.71*    \\ \hline
				Switzerland     & 95.52  & 96.33  &62.63 & 95.60  & 139.56 & 86.04*    & 95.00    \\ \hline
				Czech Republic  & 71.33  & 74.49  &87.92 & 86.08  & 74.94  & 49.01*    & 63.91    \\ \hline
				Germany         & 102.72 & 85.70  &106.71 & 83.40  & 77.96  & 42.18*    & 62.77    \\ \hline
				Denmark         & 30.74  & 40.65  &84.52 & 42.49  & 46.11  & 31.09    & 28.09*    \\ \hline
				Estonia         & 87.48  & 122.11 &244.91 & 100.38 & 117.31 & 57.70*    & 64.45    \\ \hline
				Spain           & 132.35 & 134.81 &126.32 &147.66 & 226.93 & 131.89*   & 135.57   \\ \hline
				Finland         & 21.31  & 39.76  &45.30 & 23.88  & 28.00  & 18.00    & 16.57*    \\ \hline
				France          & 22.80  & 26.10  &52.86 & 21.44  & 22.35  & 14.91*    & 20.49    \\ \hline
				UK              &149.87  &332.34  &298.71 &171.00 &176.16 &140.35 &133.75* \\ \hline
				Greece          & 72.11  & 48.01  &96.65 & 46.72  & 35.61  & 30.68*    & 43.69    \\ \hline
				Hungary         & 29.69  & 45.54  &61.60 & 45.84  & 29.52  & 16.13*    & 26.50    \\ \hline
				Ireland         & 71.46  & 70.47  &95.88 & 68.17  & 63.39  & 45.23*    & 67.66    \\ \hline
				Italy           & 120.33 & 120.51 &152.04 & 85.24  & 80.58  & 48.66*    & 111.43   \\ \hline
				Lithuania       & 86.81  & 122.01 &213.56 & 78.77  & 98.17  & 61.59*    & 76.44    \\ \hline
				Latvia          & 66.62  & 79.01  &89.68 & 62.05  & 83.67  & 49.11*   & 56.59    \\ \hline
				Montenegro      & 166.84 & 170.73 &234.48 & 132.44 & 161.84 & 97.44*    & 151.89   \\ \hline
				North Macedonia & 74.44  & 65.44  &89.26 & 82.73  & 92.14  & 67.37    & 58.81*    \\ \hline
				Netherlands     & 53.99  & 84.66  &52.94 & 56.57  & 55.30  & 43.27*    & 44.11    \\ \hline
				Norway          & 59.36  & 81.32  &167.89 & 69.09  & 57.32  & 37.70*    & 42.38    \\ \hline
				Poland          & 64.03  & 68.62  &85.79 & 62.75  & 51.59  & 50.30*    & 52.09    \\ \hline
				Portugal        & 56.50  & 56.74  &80.74 & 66.73  & 61.95  & 48.90*    & 53.75    \\ \hline
				Romania         & 37.06  & 43.55  &47.62 & 51.20  & 43.31  & 33.02*    & 36.96    \\ \hline
				Serbia          & 57.00  & 42.77  &96.06 & 71.87  & 85.83  & 44.44*    & 51.72    \\ \hline
				Sweden          & 67.97*  & 108.96 &102.01 & 71.79  & 59.68  & 69.59    & 74.19    \\ \hline
				Slovenia        & 45.96  & 40.63  &59.61 & 53.34  & 40.37  & 29.23*    & 31.55    \\ \hline
				Slovakia        & 34.92  & 36.84  &60.47 & 43.46  & 42.73  & 27.90*    & 33.69    \\ \hline
			\end{tabular}
		}
	\end{table}
	
\begin{figure}[p]
		\centering{\includegraphics[width=\textwidth]{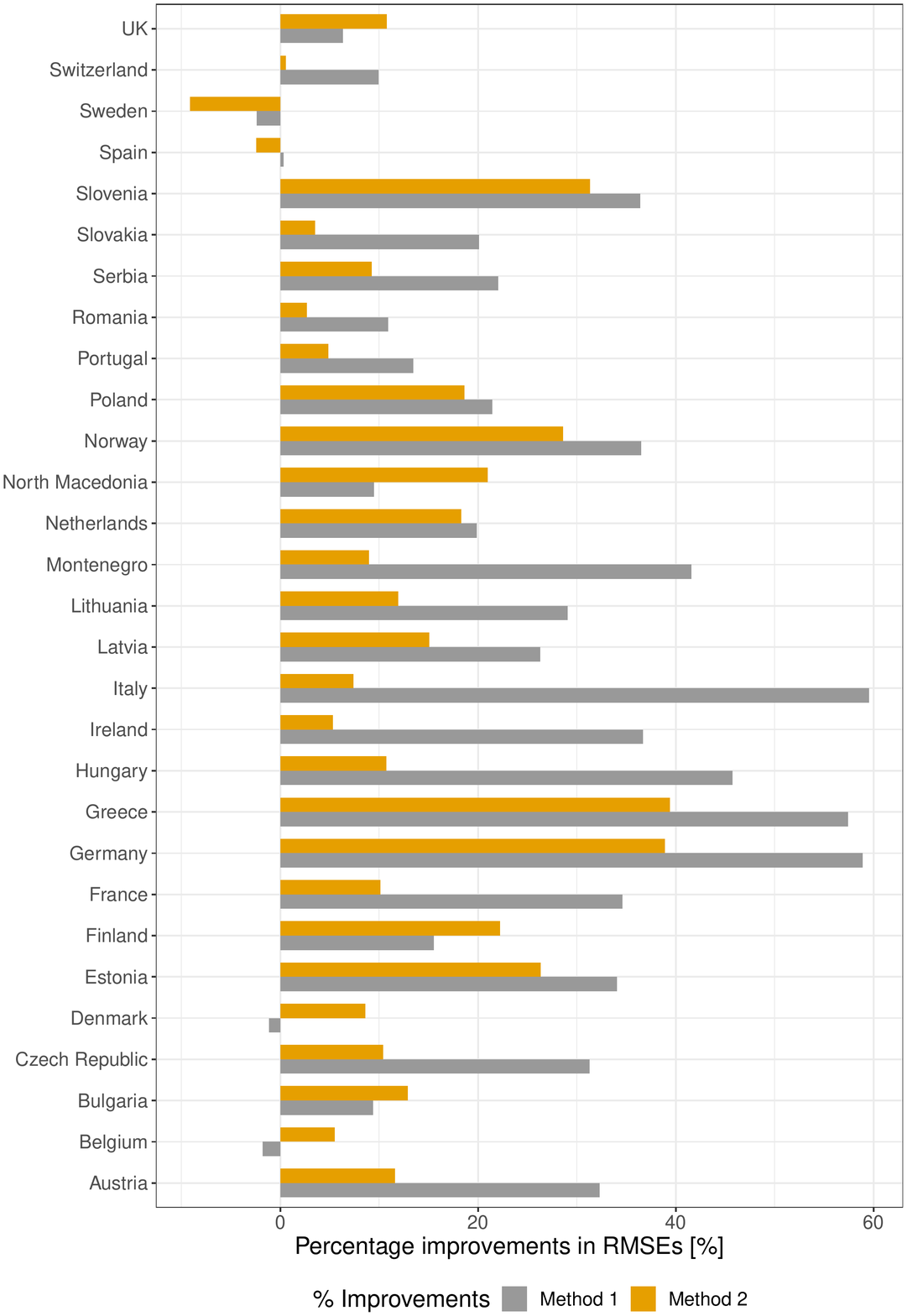}}
		\caption{Percentage improvements in RMSE with respect to ARIMA.}
		\label{err_arima}
	\end{figure}

\begin{figure}[p]
		\centering{\includegraphics[width=\textwidth]{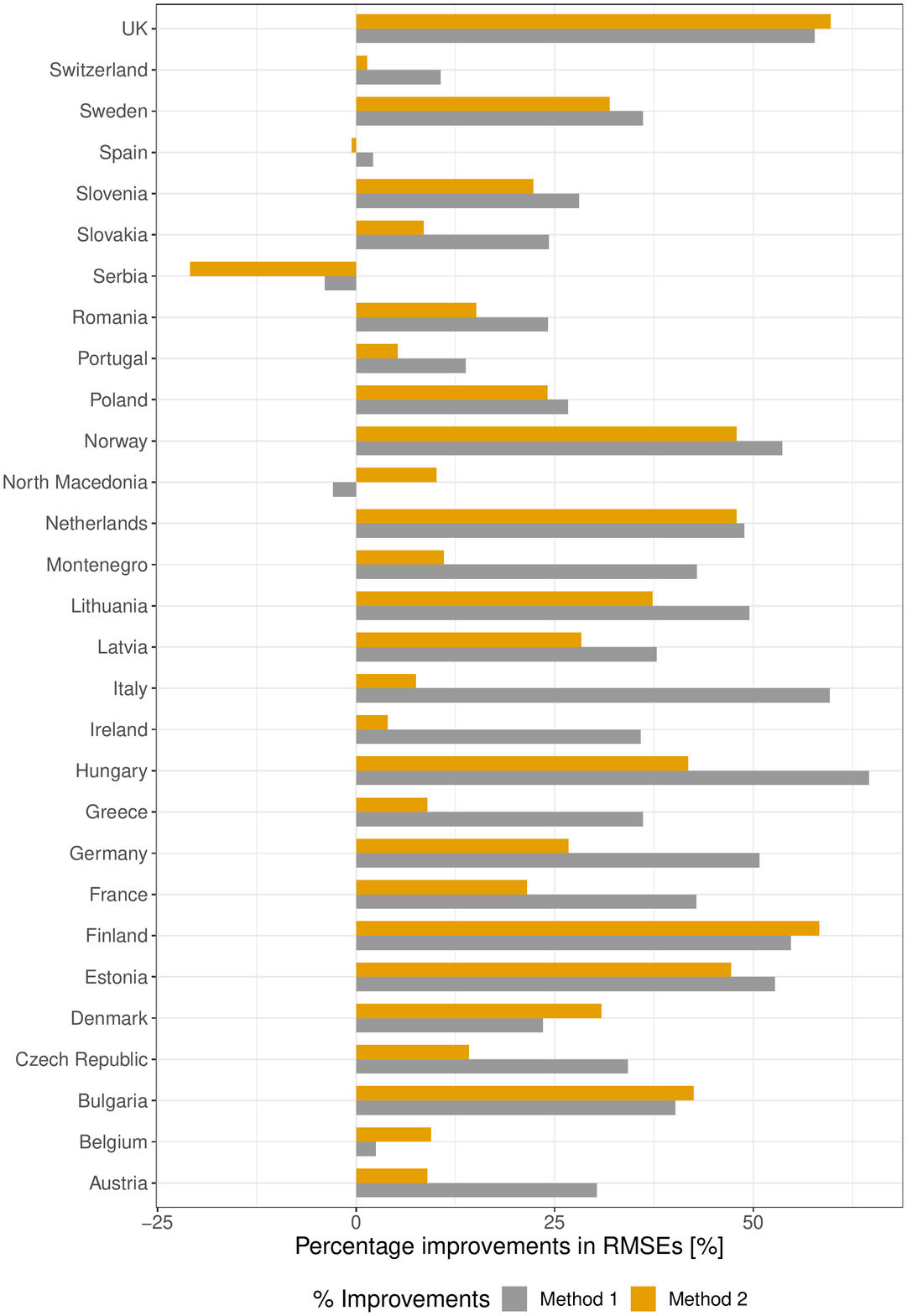}}
		\caption{Percentage improvements in RMSE with respect to FFNN.}
		\label{err_FFNN}
	\end{figure}

	Based on Tables \ref{tr1}, \ref{tr2}, and \ref{tr3}, the following points can be observed:
	
	\begin{itemize}
		\item For the sample French data set, both proposed methods have performed better than other models for 48 hours ahead CO$_2$ intensity forecasting. The RMSE, MAE{, and} MAPE of Method 1 are 15.57, 11.99{, and} 11.47 for the recursive approach; and 15.53, 11.80{, and} 11.46 for Dirrec one. Whereas, in Method 2, for the recursive approach, these errors are 18.28, 14.58{, and} 15.98; and for the Dirrec approach are 18.16, 14.43{, and} 15.91, respectively. In terms of RMSE, Method 1 has 14.82\% and 14.48\% improvement in forecasting in comparison to Method 2 for recursive and Dirrec approaches.
		
		The percentage of improvements with Method 1 as compared to all other methods in the study are shown in Table \ref{tr11} for both recursive and Dirrec approaches.
		
		\begin{table}[h]
			\centering
			\caption{Percentage improvements of Method 1 with state-of-the-art models for CO$_2$ intensity data in France.}
			\label{tr11}
			\begin{tabular}{|c|c|c|c|c|c|c|}
				\hline
				\multirow{2}{*}{Models} & \multicolumn{3}{|c|}{Recursive approach} & \multicolumn{3}{|c|}{Dirrec approach} \\ \cline{2-7}
				& RMSE        & MAE         & MAPE       & RMSE       & MAE        & MAPE      \\ \hline
				ARIMA                   & 46.04       & 47.48       & 46.99      & 45.20      & 47.34      & 46.44     \\ \hline
				FFNN                    & 41.08       & 40.82       & 45.74      & 40.10      & 40.19      & 45.11     \\ \hline
				SVM                     & 75.52       & 76.23       & 80.62      & 73.23      & 73.12      & 79.18     \\ \hline
				PSF                     & 23.93       & 21.61       & 24.83      & 23.79      & 22.82      & 24.65     \\ \hline
				DPSF                    & 40.39       & 54.28       & 49.29      & 40.38      & 46.06      & 49.20     \\ \hline
				Method 2                & 14.82       & 17.69       & 28.22      & 14.48      & 18.22      & 27.96     \\ \hline
			\end{tabular}
		\end{table}
		
		\item Dirrec is found to be an expensive forecasting strategy in which the first forecasted value is appended to the data set and a new model is fitted to forecast the next value. It means, for 48 hours ahead forecast, the execution time of the Dirrec model is nearly 48 times that of the recursive one. From Tables \ref{tr1} and \ref{tr11}, it can be seen that the error improvements with Dirrec strategy are negligible as compared to the recursive one. Hence, the recursive strategy comes out to be the better-suited forecasting strategy for short-term multi-step CO$_2$ intensity forecasting.
		
		\item Based on Table \ref{tr2}, with the Monte-Carlo strategy, among 29 European countries, Method 1 outperformed all methods for 22 countries{, and} for 6 countries, Method 2 was the best performing method. The overall average percentage improvements of Method 1 as compared to ARIMA and FFNN methods range from -2.38\% (for Sweden) to 58.93\% (for Germany) and -3.97\% (for Serbia) to 64.58\% (for Hungary), respectively in terms of RMSE collectively for all countries. Similarly, the same improvements ranges with Method 2 are -9.15\% (for Sweden) to 39.41\% (for Greece) and -20.92\% (for Serbia) to 59.74\% (for UK), respectively. Here, the negative sign indicates that Method 1 or 2 performed in a less efficient way than the methods in the comparison study. The country-wise percentage improvements in RMSE for ARIMA and FFNN methods are shown in Figures \ref{err_arima} and \ref{err_FFNN}, respectively.
		
		\item Method 1 achieved superior accuracy with the statistical significance at the cost of computational complexity. Considering the computational complexity (especially, the time complexity), Method 1 ($O(N^2)$) observed more complex than Method 2 ($O(N log N)$).
		
	\end{itemize}
	
	\newpage
	\section{Impact on electricity market scheduling}
\begin{figure}[h]
		\centering{\includegraphics[width=\textwidth]{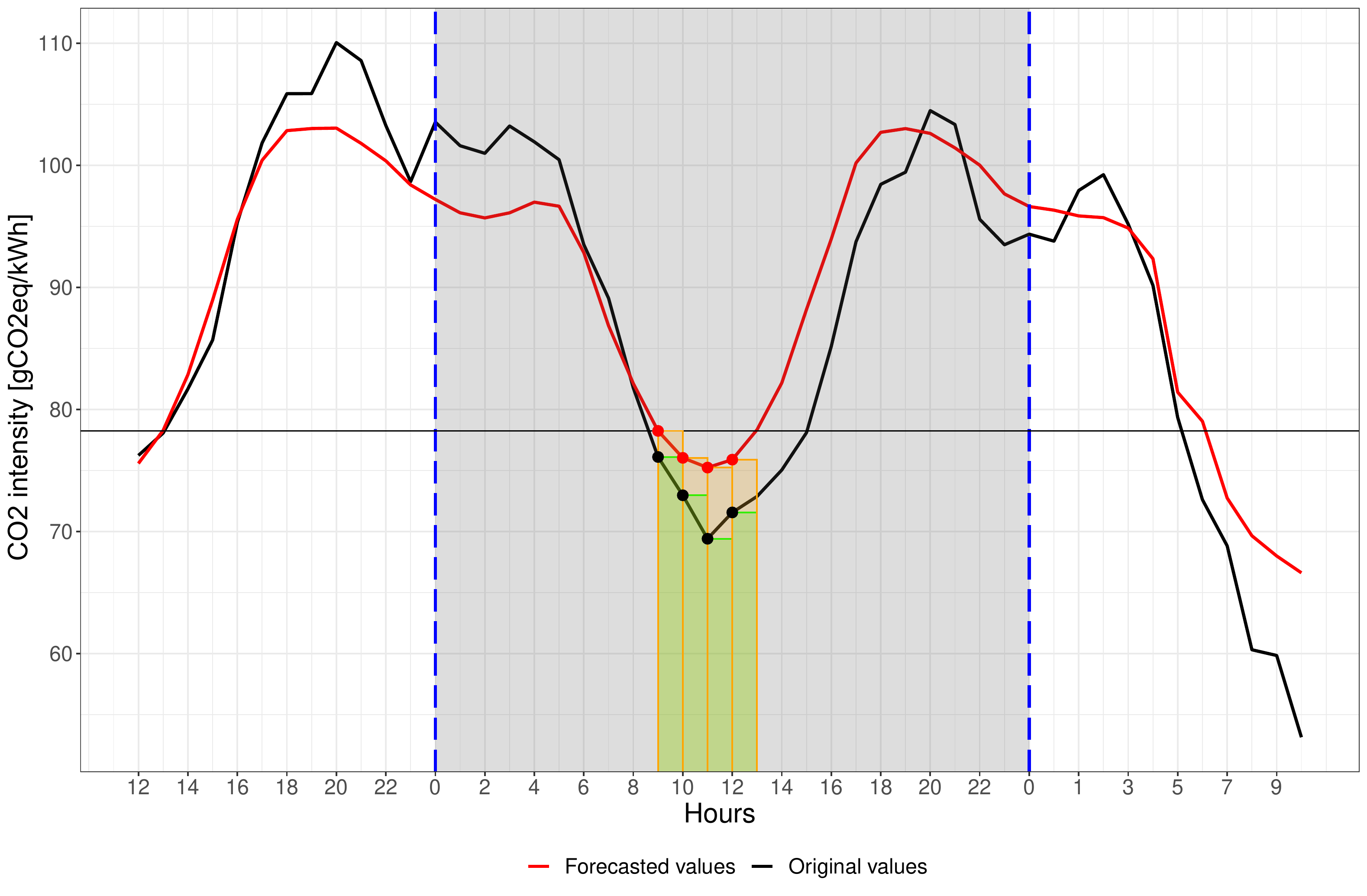}}
		\caption{Example of scheduling 4 hours of flexible electricity consumption one day in advance in France. The gray area shows the day-ahead interval. The red and black lines show forecasted and realized values of the CO$_2$ intensity, respectively. The four hours minimum of the forecast is market with red dots. The corresponding realized values are marked with black dots.}
		\label{hour_4}
	\end{figure}

	The ultimate goal of the short-term CO$_2$ intensity forecast is to enable {the} scheduling of flexible demand in the electricity market to minimize the resulting CO$_2$ emissions. Figure \ref{hour_4} shows the realized CO$_2$ emissions {per kWh} electricity consumed (black) and corresponding forecasted values (red) with Method 1 for France for consecutive 48 hours. The start point of the time series (12.00 PM) is the point in time today (day $D$) at which the day-ahead electricity market is cleared. Bids submitted to the day-ahead power exchange before this time is targeted the whole of the next-coming day (24 hours of {the} day $D+1$). This interval is marked in {the} grey between the dashed blue lines. 
	
	In this example, a flexible electricity consumer is willing to bid in specific hours in day $D+1$ to minimize the associated CO$_2$ emissions. Specifically, electricity consumption {is scheduled} during the four hours with least forecasted CO$_2$ intensity. The dots on the red line denote the four minimum values of the forecast within the 24-hour window. The horizontal line shows the maximum CO$_2$ emissions {per} kilowatt-hour electricity consumed value within the scheduled hours with reference to the forecasted series. The dots on the black line highlight the corresponding realized values. The mean intensity for these four hours is 76.35 gCO2eq/kWh and 72.51 gCO2eq/kWh, respectively, for the forecasted and realized values.
	
	Similar examples of scheduling for Germany, Norway, Denmark{, and} Poland are shown in Figure \ref{fig:four graphs}. These countries have been selected based on their location in Figure~\ref{fig6x}: Poland with {a} very low share of non-fossil generation, France with {a} high share of nuclear, Norway with {a} high share of hydro, Denmark with {a} high share of wind and Germany somewhere in between.
	
	%
	%
	%
	
\begin{figure}[h]
		\centering{\includegraphics[width=\textwidth]{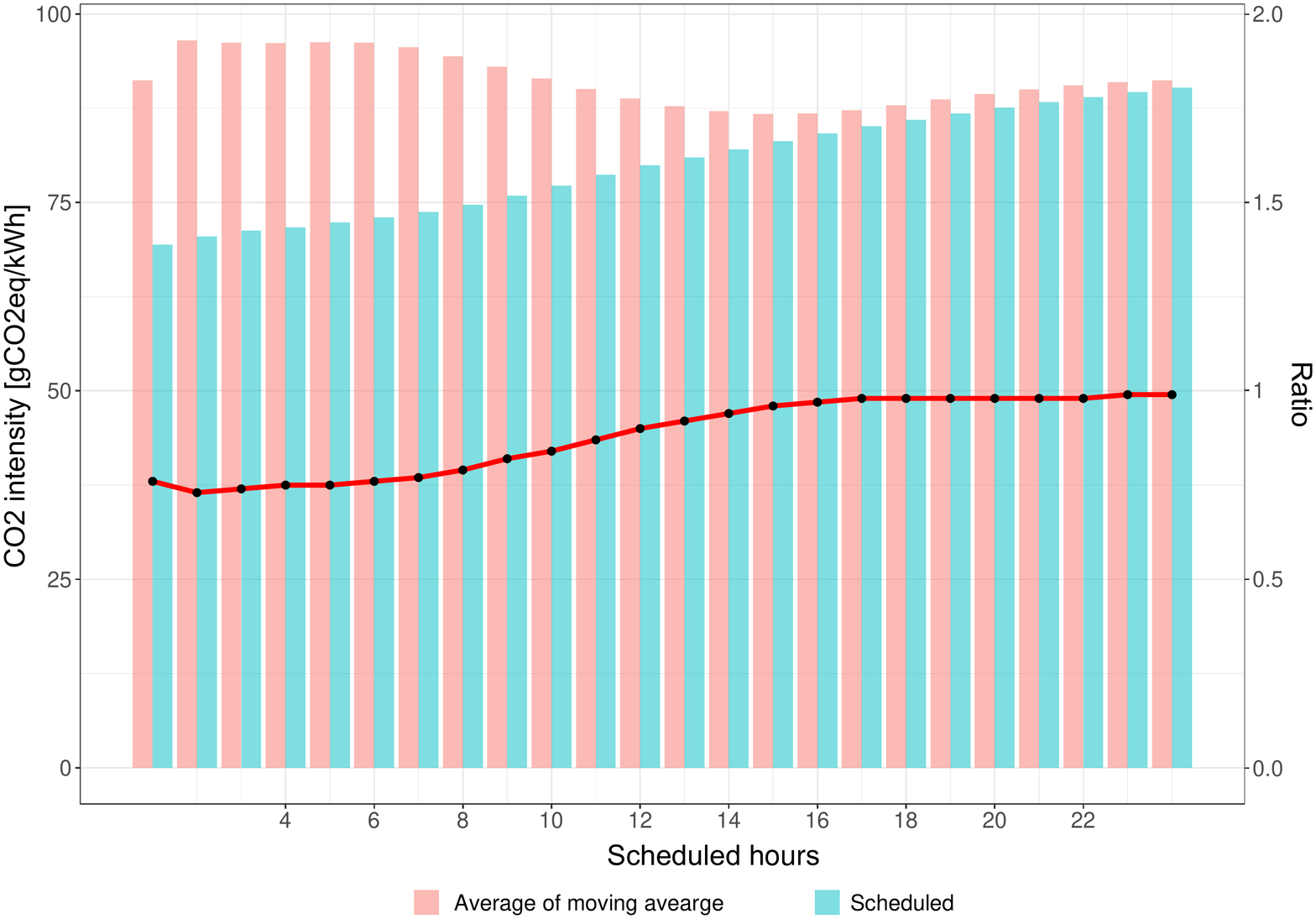}}
		\caption{Comparison of emissions between scheduling and consuming at a random time during the 24-hour interval for an entire year in France. Each bar shows results for different durations of flexible consumption. The ratio between the bars is shown as the red line.}
		\label{Overall}
	\end{figure}

	Figure \ref{hour_4} shows just a single example of scheduling four hours of electricity consumption in France. In Figure \ref{Overall}, the emissions between scheduling and consuming {are compared} at a random time during the 24-hour interval for an entire year in France. Each bar shows results for different durations of flexible consumption from one to 24 hours. The ratio between the bars is shown as the red line. A ratio of one means there are no savings from scheduling compared to consuming at a random time. This is the case when the flexible consumption takes up all 24 hours of the window. The ratio decreases as the duration of flexible consumption decreases. This is to be expected as {a} shorter duration of consumption means more flexibility for scheduling during the 24 {windows}.
	
\begin{figure}[h]
		\centering{\includegraphics[width=\textwidth]{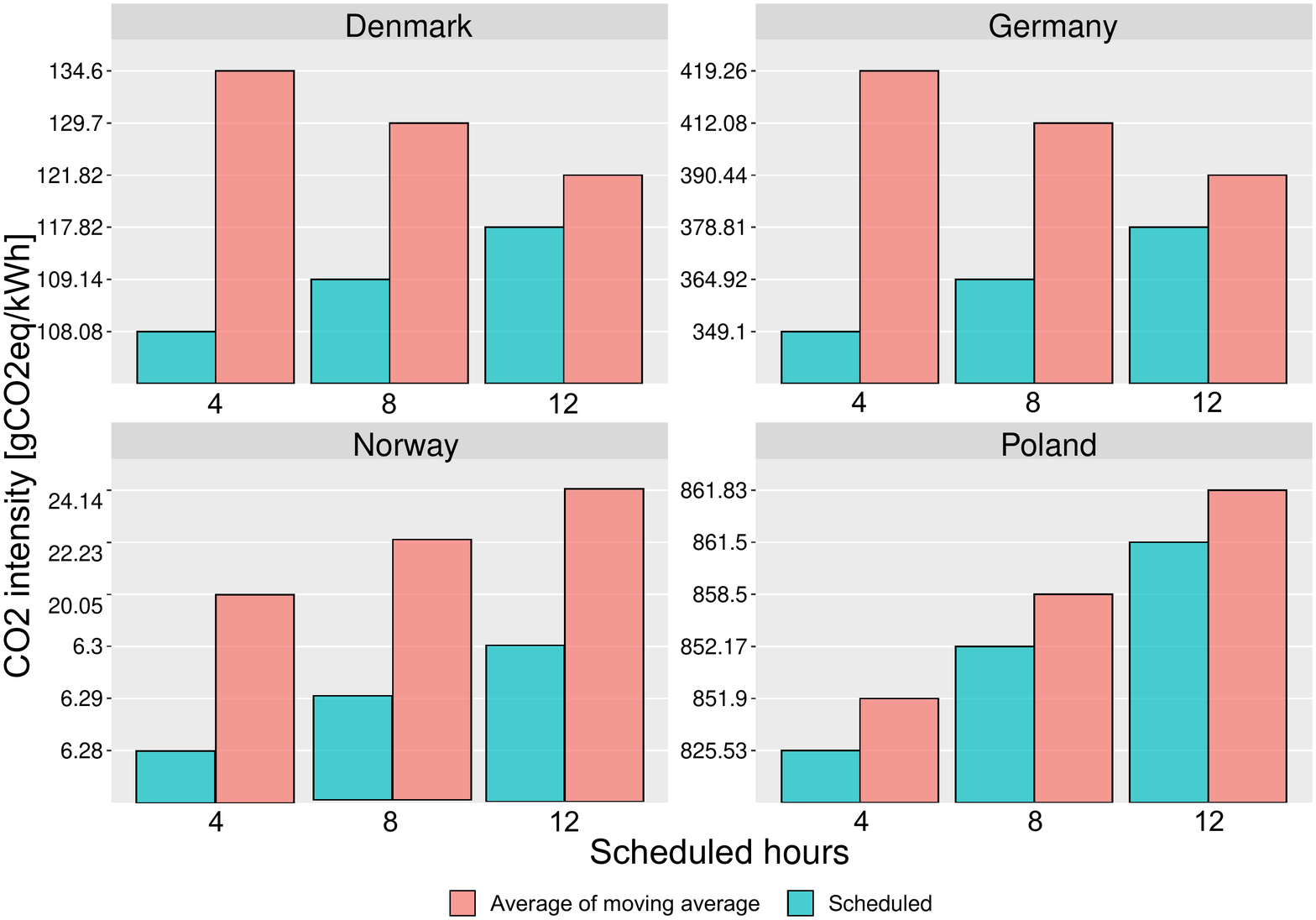}}
		\caption{Comparison of emissions between scheduling and consuming at a random time during the 24-hour interval for an entire year in four countries. The bars show results for durations of flexible consumption of 4, 8 and 12 hours.}
		\label{DGNP}
	\end{figure}

	Figure \ref{DGNP} shows similar results for Germany, Norway, Denmark{, and} Poland. Comparing the benefit of scheduling between the five countries, for a duration of flexible consumption of four hours, {observed} on average 25 \% emissions reductions in France, 17 \% in Germany, 69 \% in Norway, 20 \% in Denmark, and just 3 \% in Poland. 
	
\begin{figure}
		\centering{\includegraphics[width=\textwidth]{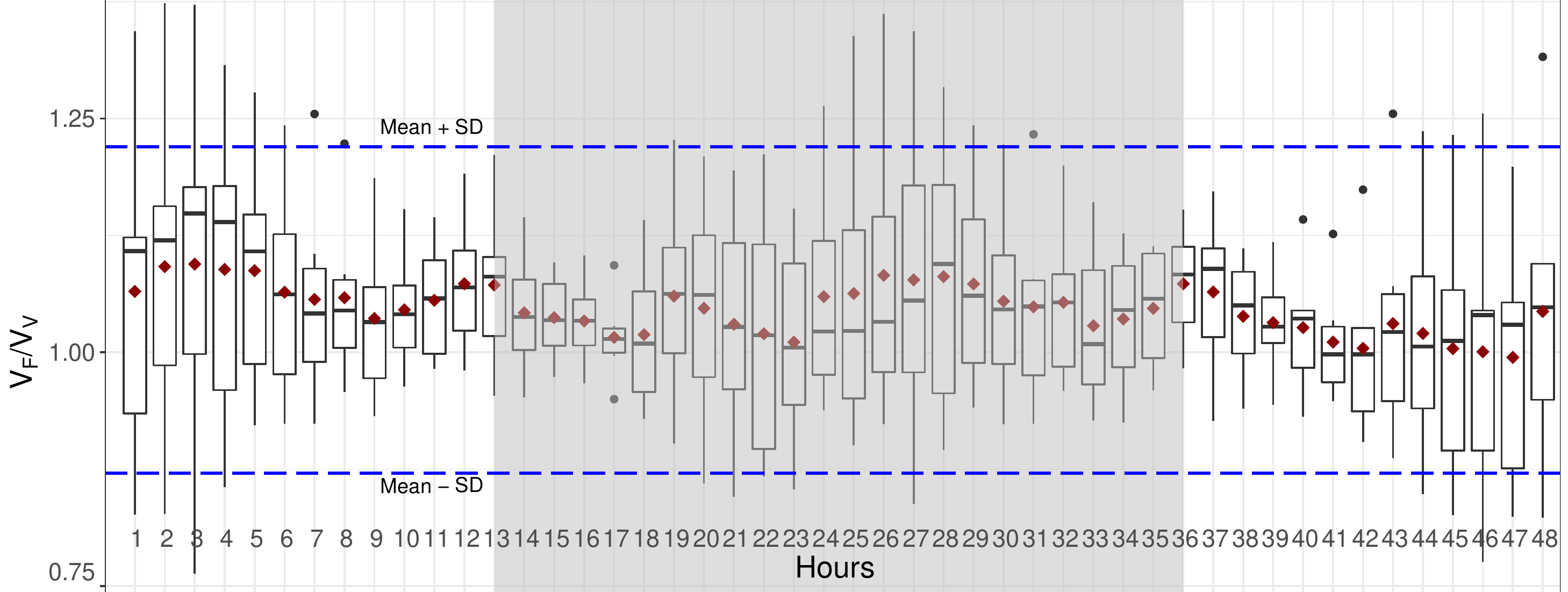}}
		\caption{Variation in the ratio of forecasted ($V_F$) and realized ($V_V$) values for each of the 48 hours of the forecast horizon. The red dots in are the mean values of these ratios for the whole year. The dashed blue lines show the standard deviations from the mean values throughout the year. The shaded regions show the 24 hours of the forecasts, which are to be used for the scheduling of flexible consumption.}
		\label{France_MC}
	\end{figure}

	Furthermore, Figure \ref{France_MC} shows the variation in the ratio of forecasted ($V_F$) and realized ($V_V$) values for each of the 48 hours of the forecast horizon with Monte-Carlo strategy with 50 iterations for France. This figure represents the statistical relation between forecasted and realized values in terms of box plots. {The red dots are} the mean values of these ratios for the whole year. The dashed blue lines show the standard deviations from the mean values throughout the year. The shaded regions in the figure show the 24 hours of the forecasts, which are to be used for the scheduling of flexible consumption. It is observed that the forecast at initial and last 8 hours {is} more consistent throughout the year, whereas, those in middle hours (day-time hours) express higher variations. This outcome concludes that, though these variations are not much significant (merely varies between 0.091 to 0.185) throughout the year, the electricity market bidder could consider a bit higher chances of deviation in the day time hour values. Similar results for Denmark, Germany, Norway, and Poland are shown in Figure~\ref{fig:four graphs_MC}.
	
	
	
	\section{Conclusions}
	Accurately forecasting the CO$_2$ intensity plays a crucial role in the socio-economic and environmental benefits derived from appropriate energy management. It can affect the energy policies of a country to a certain extent. Hence, more accurate methods to improve CO$_2$ intensity forecasts become sensible. A simple model without preprocessing of a chaotic time series usually seems {unachievable} in most of the cases. Hence, {the} preprocessing of the data set with suitable methodologies enhances the forecasting accuracy. Preprocessing with decomposition methods helped many researchers in superior forecasting outcomes.
	
	In this study, the {accuracies} of two decomposition strategies {are} examined for 48 hours ahead of CO$_2$ intensity data forecasting. The first decomposition approach is a pure and simple statistical method that decomposes a time series into seasonal, trend{, and} random series components. Similarly, the second decomposition approach bases on the EEMD method again decomposes a time series into three components, those are, high-frequency, low-frequency{, and} trend series components. These decomposition approaches separate outs of distinct meaningful patterns from the original time series and assists in achieving forecast-friendly data series components. Summarizing the observation, it can be concluded that the performance of meaningful decomposition approaches can improve the multi-step short-term CO$_2$ intensity forecast as discussed in the paper. {In this study, for CO$_2$ intensity forecasting, the first decomposition-based method (with time complexity $O(N^2)$) has been found to outperform the second method with time complexity $O(N log N)$). For 23 of the 29 European countries, the first decomposition method obtained the highest accuracy. The result is a trade-off between forecast accuracy and computational complexity. The memory complexity of both methods is the same, i.e., $O(log N)$. In this case, the higher computing time is acceptable since the forecast is only performed once per day for the day-ahead market.}
	
	After validating the {correctness} of Method 1 of the proposed forecasting method, it was used to assess the impact scheduling of flexible electricity consumption. The CO$_2$ emissions from scheduling were studied for France, Germany, Norway, Denmark{, and} Poland. It was found for all countries that scheduling the consumption after the forecasted intensity compared with consuming at a random time leads to reductions in emissions. Reductions increase with decreasing duration of consumption because of {the} increased flexibility of scheduling within the 24-hour window. In the case of a duration of flexible consumption of four hours, {it} found emissions reductions ranging from 3 \% in Poland to 69 \% in Norway.
	
	In future work, forecasts with a longer horizon and finding the optimal trade-off between CO$_2$ emissions and electricity prices can be examined to propose a meaningful strategy for scheduling flexible electricity consumption in the day-ahead and future markets of electricity.
	
	
	\section*{Acknowledgments}
	This study was funded by Apple Inc. as part of the APPLAUSE bio-energy collaboration with Aarhus
	University. Bo Tranberg acknowledges funding from ELFORSK via project 351-054.

	\appendix
	\renewcommand\appendixname{}
	\counterwithin{table}{section}
	\counterwithin{figure}{section}
	
	\section{Tables}
	
	\begin{landscape}
		\begin{table}[]
			\centering
			\caption{Complexity study of the proposed methods.}
			\label{tr3}
			\begin{tabular}{|c|c|c|c|c|c|}
				\hline
				\multicolumn{6}{|c|}{Complexity of Method 1}                                                                                                                                                                                                                                                                                                                                                                                                                       \\ \hline
				\multicolumn{1}{|c|}{Complexity} & \begin{tabular}[c]{@{}c@{}}Decomposition with\\ moving averages\end{tabular}  & \begin{tabular}[c]{@{}c@{}}Seasonal component\\ forecast with FFNN\end{tabular}    & \begin{tabular}[c]{@{}c@{}}Random component\\ forecast with ARIMA\end{tabular}    & \multicolumn{1}{c|}{\begin{tabular}[c]{@{}c@{}}Trend component\\ forecast with ARIMA\end{tabular}} & Overall                                                           \\ \hline
				\multicolumn{1}{|c|}{Time}       & \begin{tabular}[c]{@{}c@{}}Constant\\ $O(1)$\end{tabular}                      & \begin{tabular}[c]{@{}c@{}}Quadratic\\ $O(N^2)$\end{tabular}                          & \begin{tabular}[c]{@{}c@{}}Constant\\ $O(1)$\end{tabular}                           & \multicolumn{1}{c|}{\begin{tabular}[c]{@{}c@{}}Logarithmic\\ $O(log N)$\end{tabular}}                & \begin{tabular}[c]{@{}c@{}}Quadratic\\ $O(N^2)$\end{tabular}         \\ \hline
				\multicolumn{1}{|c|}{Memory}     & \begin{tabular}[c]{@{}c@{}}Logarithmic\\ $O(log N)$\end{tabular}               & \begin{tabular}[c]{@{}c@{}}Constant\\ $O(1)$\end{tabular}                            & \begin{tabular}[c]{@{}c@{}}Logarithmic\\ $O(log N)$\end{tabular}                    & \multicolumn{1}{c|}{\begin{tabular}[c]{@{}c@{}}Logarithmic\\ $O(log N)$\end{tabular}}                & \begin{tabular}[c]{@{}c@{}}Logarithmic\\ $O(log N)$\end{tabular}    \\ \hline
				\multicolumn{6}{|c|}{Complexity of Method 2}                                                                                                                                                                                                                                                                                                                                                                                                                       \\ \hline
				\multicolumn{1}{|c|}{Complexity} & \begin{tabular}[c]{@{}c@{}}Decomposition with\\ EEMD method\end{tabular} & \begin{tabular}[c]{@{}c@{}}High-freq. component\\ forecast with ARIMA\end{tabular} & \begin{tabular}[c]{@{}c@{}}Low-freq. component\\ forecast with ARIMA\end{tabular} & \multicolumn{1}{c|}{\begin{tabular}[c]{@{}c@{}}Trend component\\ forecast with ARIMA\end{tabular}} & Overall                                                           \\ \hline
				\multicolumn{1}{|c|}{Time}       & \begin{tabular}[c]{@{}c@{}}Linearithmic\\ $O(N log N)$\end{tabular}            & \begin{tabular}[c]{@{}c@{}}Constant\\ $O(1)$\end{tabular}                            & \begin{tabular}[c]{@{}c@{}}Constant\\ $O(1)$\end{tabular}                           & \multicolumn{1}{c|}{\begin{tabular}[c]{@{}c@{}}Constant\\ $O(1)$\end{tabular}}                       & \begin{tabular}[c]{@{}c@{}}Linearithmic\\ $O(N log N)$\end{tabular} \\ \hline
				\multicolumn{1}{|c|}{Memory}     & \begin{tabular}[c]{@{}c@{}}Logarithmic\\ $O(log N)$\end{tabular}               & \begin{tabular}[c]{@{}c@{}}Constant\\ $O(1)$\end{tabular}                            & \begin{tabular}[c]{@{}c@{}}Constant\\ $O(1)$\end{tabular}                           & \multicolumn{1}{c|}{\begin{tabular}[c]{@{}c@{}}Logarithmic\\ $O(log N)$\end{tabular}}                & \begin{tabular}[c]{@{}c@{}}Logarithmic\\ $O(log N)$\end{tabular}    \\ \hline
			\end{tabular}
		\end{table}
	\end{landscape}
	
	\newgeometry{textwidth=7in, textheight=9in}
	\section{Figures}
	
	\begin{figure}[h]
		\centering
		\begin{subfigure}[b]{0.49\textwidth}
			\centering{\includegraphics[width=\textwidth]{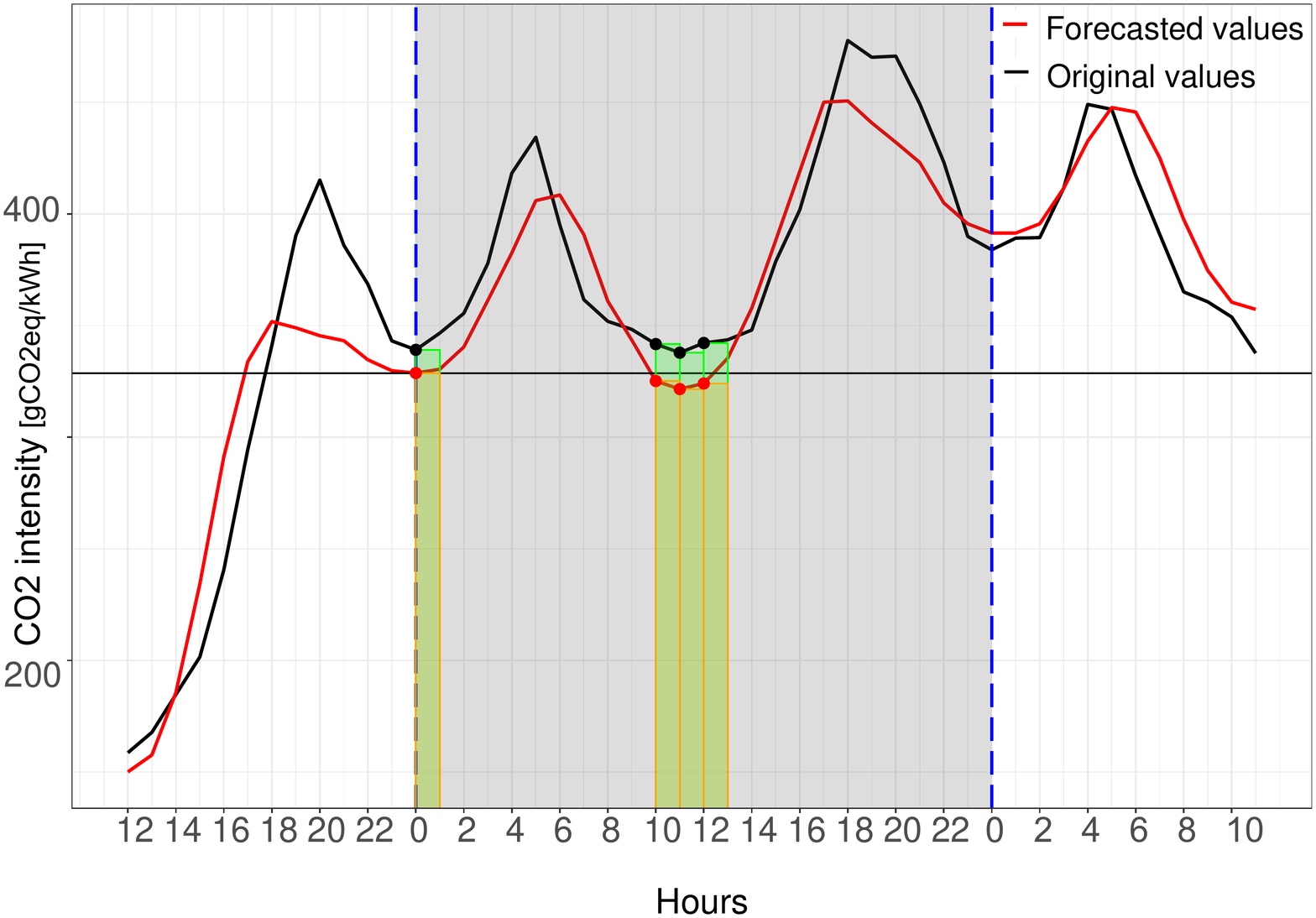}}
			\caption{Germany}
			\label{Germany}
		\end{subfigure}
		\begin{subfigure}[b]{0.49\textwidth}
			\centering{\includegraphics[width=\textwidth]{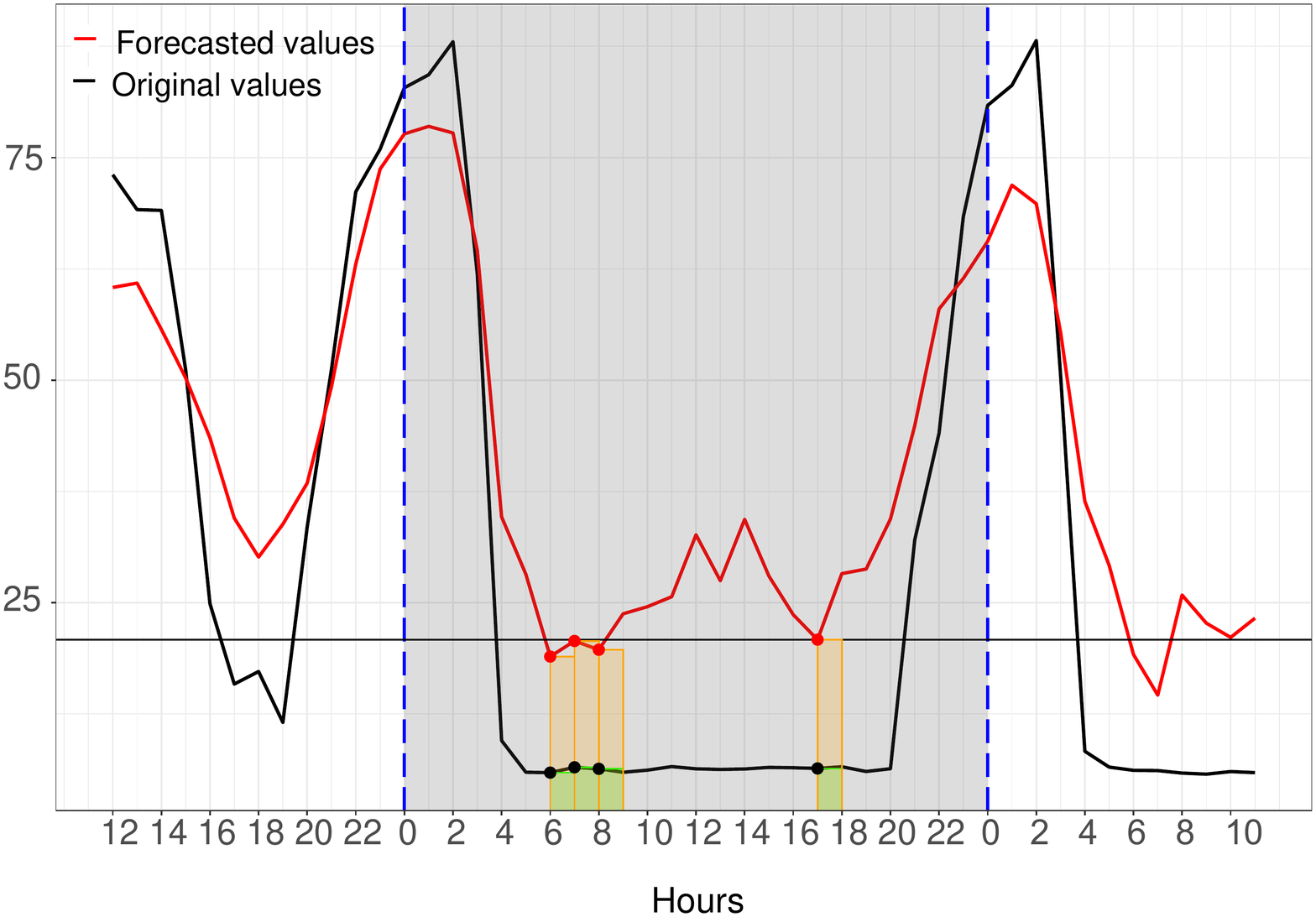}}
			\caption{Norway}
			\label{Norway}
		\end{subfigure}
		\par\bigskip
		\begin{subfigure}[b]{0.49\textwidth}
			\centering{\includegraphics[width=\textwidth]{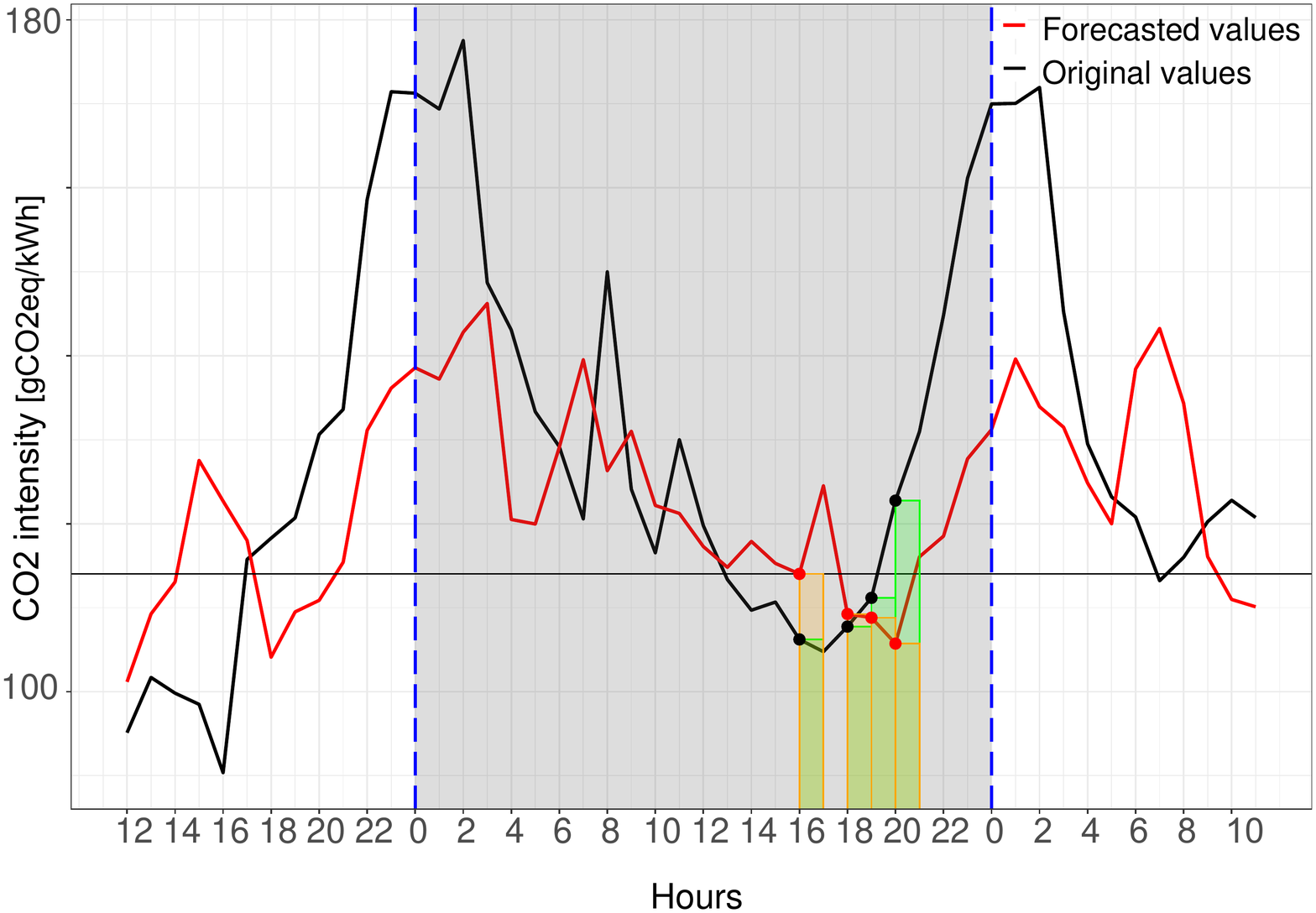}}
			\caption{Denmark}
			\label{Denmark}
		\end{subfigure}
		\begin{subfigure}[b]{0.49\textwidth}
			\centering{\includegraphics[width=\textwidth]{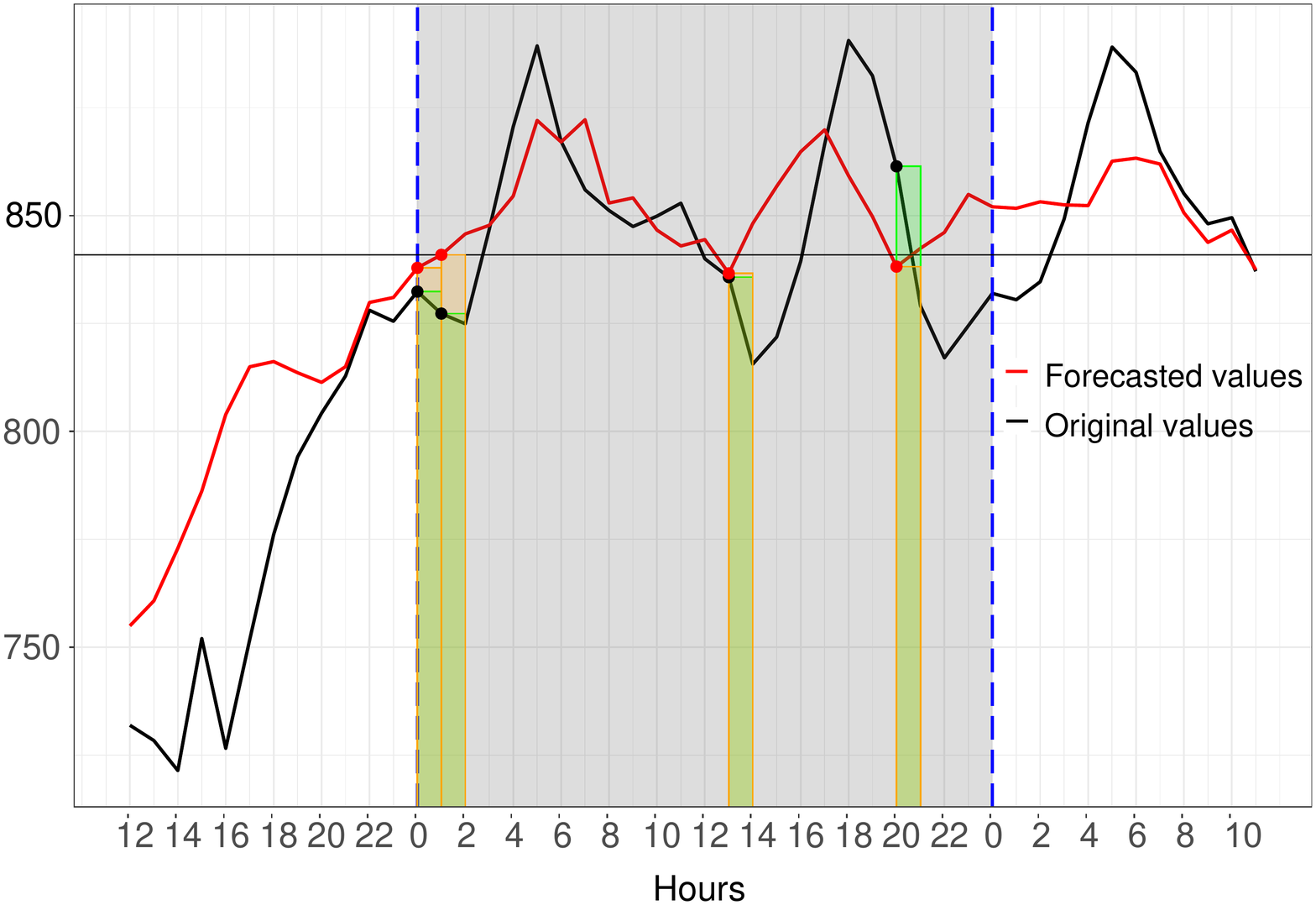}}
			\caption{Poland}
			\label{Poland}
		\end{subfigure}
		\caption{Example of scheduling 4 hours of flexible electricity consumption one day in advance in four different countries. The gray area shows the day-ahead interval. The red and black lines show forecasted and realized values of the CO$_2$ intensity, respectively. The four hours minimum of the forecast is market with red dots. The corresponding realized values are marked with black dots.}
		\label{fig:four graphs}
	\end{figure}
	\restoregeometry
	
	\newgeometry{textwidth=7in, textheight=9in}
	\begin{figure}[p]
		\centering
		\begin{subfigure}[b]{.49\textwidth}
			\centering{\includegraphics[width=\textwidth]{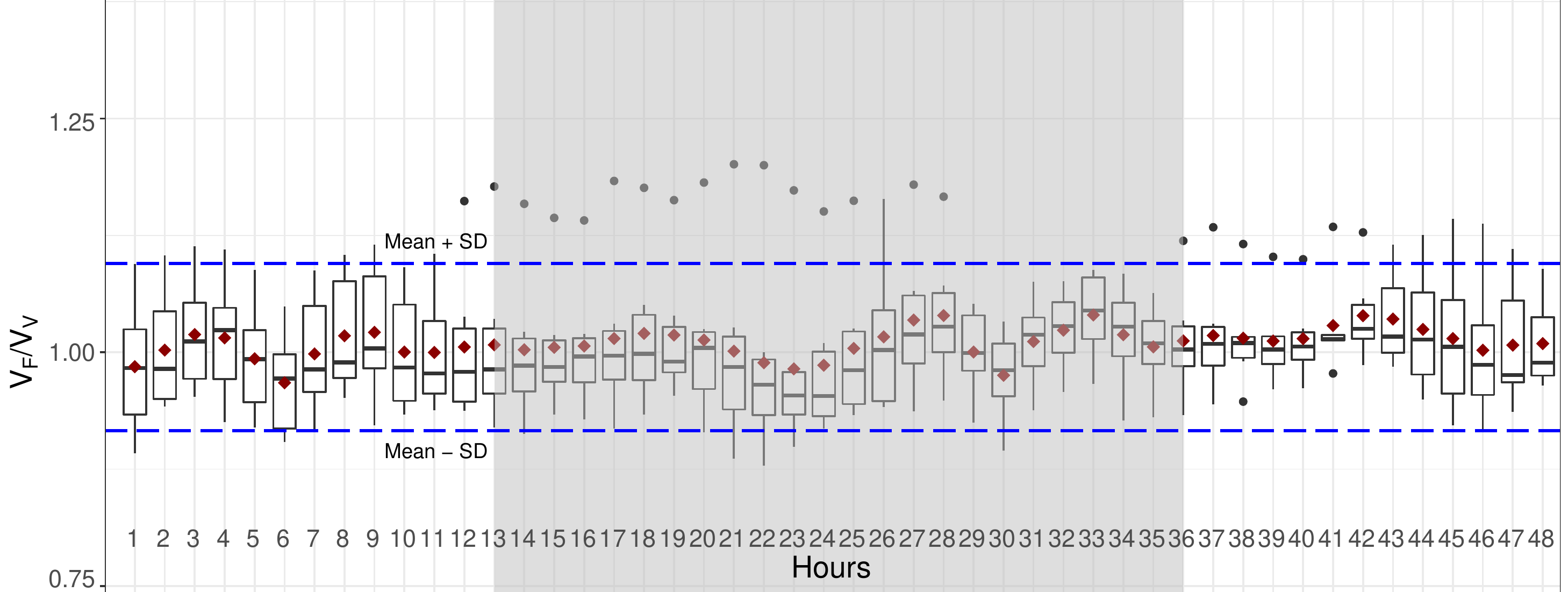}}
			\caption{Germany}
			\label{Germany_MC}
		\end{subfigure}
		\begin{subfigure}[b]{.49\textwidth}
			\centering{\includegraphics[width=\textwidth]{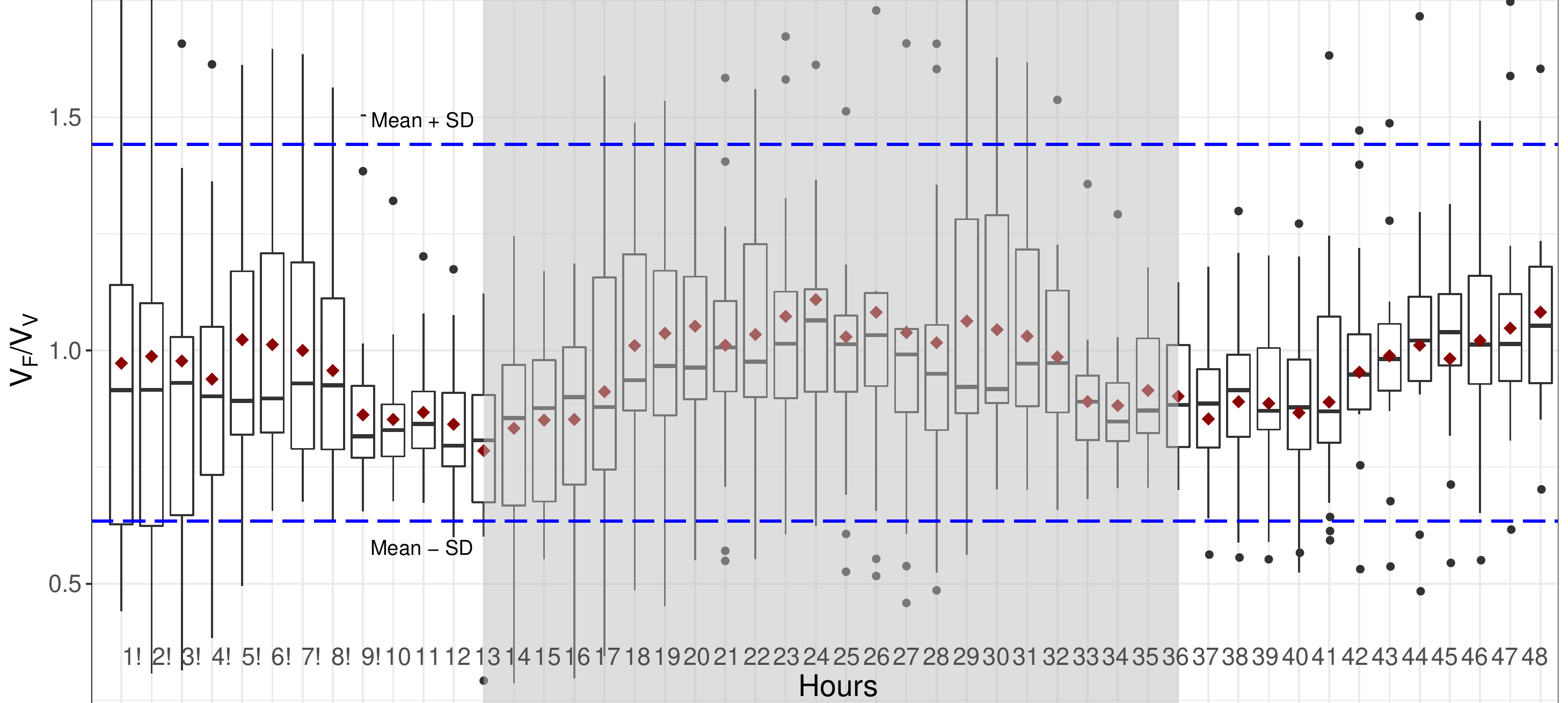}}
			\caption{Norway}
			\label{Norway_MC}
		\end{subfigure}
		\begin{subfigure}[b]{.49\textwidth}
			\centering{\includegraphics[width=\textwidth]{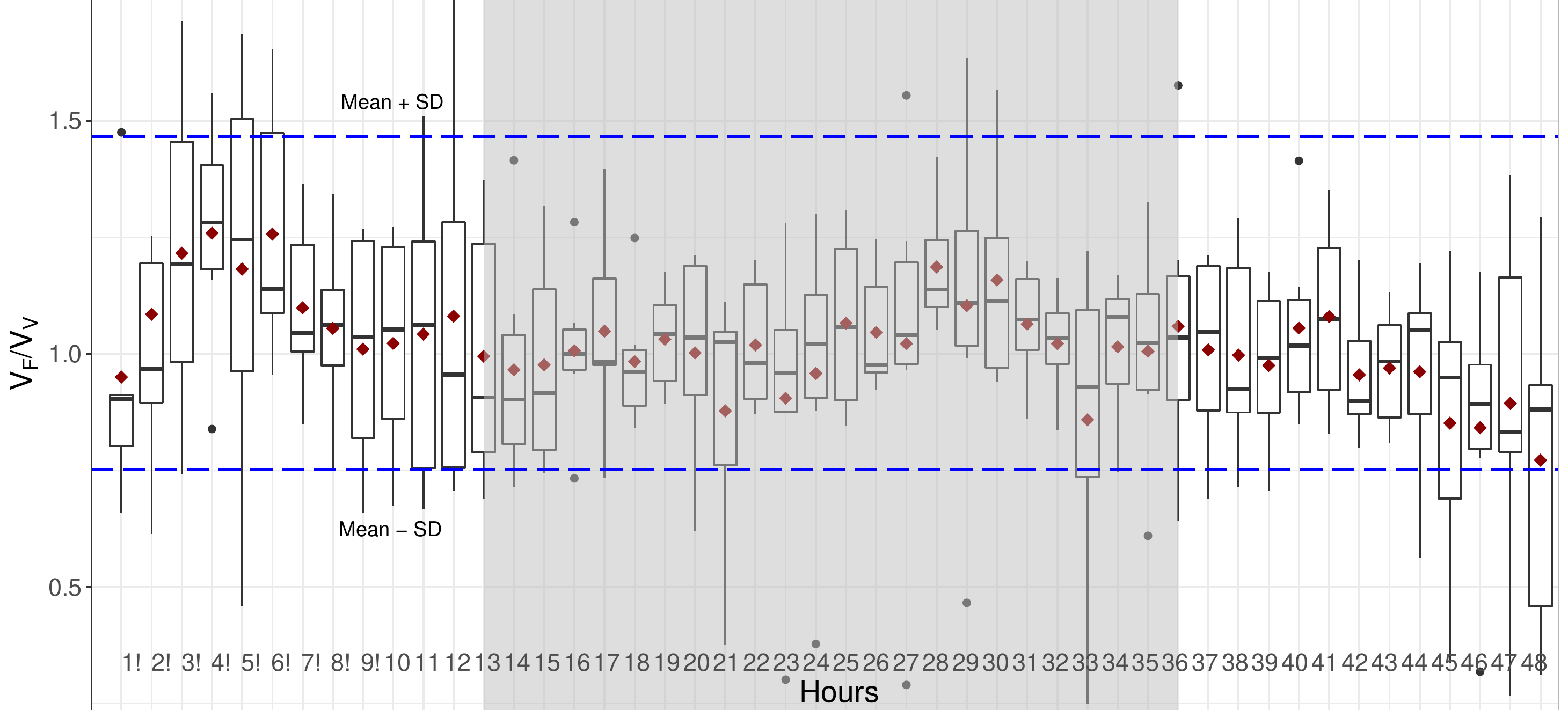}}
			\caption{Denmark}
			\label{Denmark_MC}
		\end{subfigure}
		\begin{subfigure}[b]{.49\textwidth}
			\centering{\includegraphics[width=\textwidth]{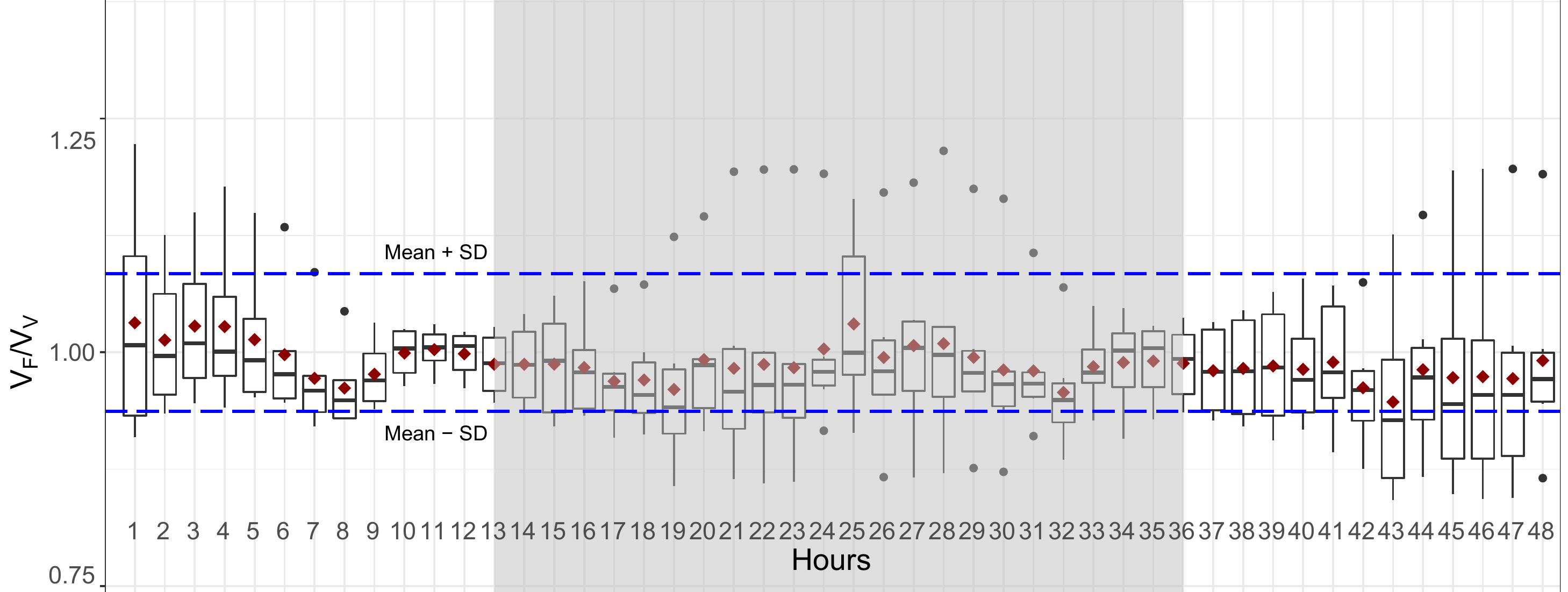}}
			\caption{Poland}
			\label{Poland_MC}
		\end{subfigure}
		\caption{Variation in the ratio of forecasted ($V_F$) and realized ($V_V$) values for each of the 48 hours of the forecast horizon. The red dots in are the mean values of these ratios for the whole year. The dashed blue lines show the standard deviations from the mean values throughout the year. The shaded regions show the 24 hours of the forecasts, which are to be used for the scheduling of flexible consumption.}
		\label{fig:four graphs_MC}
	\end{figure}
	\restoregeometry
	
	\clearpage
	\section*{References}
	
	\bibliography{mybibfile}
	
\end{document}